\documentclass[preprint,aps,prd,showpacs,floatfix]{revtex4-1}
\usepackage{times}
\usepackage {euscript}
\usepackage{amsthm}
\usepackage{graphicx}
\usepackage{amsmath}
\usepackage{amssymb}
\usepackage{amsfonts}
\usepackage[english]{babel}
\usepackage{epstopdf}
\usepackage{multirow,makecell,array}
\usepackage{mathtools}
\usepackage{epstopdf}
\usepackage{color}
\begin{document}
\thispagestyle{empty}

    \title{Dynamically assisted Schwinger effect beyond\\ the spatially-uniform-field approximation}

\author{I.~A.~Aleksandrov$^{1, 2}$}\author{G.~Plunien$^{3}$} \author{V.~M.~Shabaev$^{1}$}
\affiliation{$^1$~Department of Physics, St. Petersburg State University, 7/9 Universitetskaya Naberezhnaya, Saint Petersburg 199034, Russia\\$^2$~NRC ``Kurchatov Institute'' --- ITEP, Moscow 117218, Russia\\$^3$~Institut f\"ur Theoretische Physik, Technische Universit\"at Dresden, Mommsenstrasse 13, Dresden D-01062, Germany
\vspace{10mm}
}

\begin{abstract}
We investigate the phenomenon of electron-positron pair production from vacuum in the presence of a strong electric field superimposed by a weak but fast varying pulse which substantially increases the total particle yield. We employ a nonperturbative numerical technique and perform the calculations beyond the spatially-uniform-field approximation, i.e. dipole approximation, taking into account the coordinate dependence of the fast component. The analysis of the main characteristics of the pair-production process (momentum spectra of particles and total amount of pairs) reveals a number of important features which are absent within the previously used approximation. In particular, the structure of the momentum distribution is modified both qualitatively and quantitatively, and the total number of pairs created as well as the enhancement factor due to dynamical assistance become significantly smaller.
\end{abstract}

\maketitle
\section{Introduction}\label{sec:intro}
\indent The process of the vacuum decay accompanied by the production of electron-positron pairs in the presence of strong external fields was predicted decades ago~\cite{sauter_1931, euler_heisenberg,schwinger_1951} and still remains a very intriguing phenomenon. From the theoretical viewpoint, the interest in this effect is due to the nonperturbative nature of the pair-production process taking place in strong quasistatic backgrounds. In order to probe the quantum vacuum in this regime, i.e. to study the Schwinger mechanism, one has to employ nonperturbative evaluation methods instead of using perturbation theory, which is not applicable in this strong-coupling domain. The essential point is that the Schwinger effect has never been observed experimentally as the required field strength is extremely large. In the case of a static and spatially uniform electric field, the characteristic critical field strength is $E_\text{c} = m^2 c^3/(|e| \hbar) \approx 1.3 \times 10^{16}$~V/cm which is $3$-$4$ orders of magnitude larger than the peak electric field strength reached in modern laser pulses. Nevertheless, the laser technologies develop very rapidly, so one may expect the Schwinger mechanism to become experimentally accessible in the not too distant future. To theoretically support these studies, it is necessary to find the most promising scenarios that can be implemented in experiments.

\indent One of the possible schemes was proposed a decade ago in Ref.~\cite{schutzhold_prl_2008}. The configuration involves two laser pulses of different intensity and frequency. While the first pulse is strong and slowly varying, the second one is weak and fast. Let $E$ ($\varepsilon$) and $\Omega$ ($\omega$) be the peak strength and frequency of the strong (weak) pulse. If one introduces the Keldysh parameters $\gamma_E = m c \Omega/|eE|$ and $\gamma_\varepsilon = m c \omega/|e\varepsilon|$~\cite{keldysh}, they should satisfy $\gamma_E \ll 1$ and $\gamma_\varepsilon \gg 1$. This means that the strong pulse alone acts in the nonperturbative (Schwinger) regime whereas the individual weak pulse can be treated in the framework of perturbation theory. It turns out that the combination of these two pulses can lead to a dramatic enhancement of the particle yield. This phenomenon was first studied in Ref.~\cite{schutzhold_prl_2008}, where the external field was represented as a sum of two spatially uniform Sauter pulses without a subcycle structure (see also Refs.~\cite{orthaber_plb_2011, fey_prd_2012, kohlfuerst_prd_2013, linder_prd_2015}). The carrier of the laser pulses was taken into account in a number of subsequent studies~\cite{akal_prd_2014, hebenstreit_plb_2014, otto_plb_2015, otto_prd_2015, panferov_epjd_2016, otto_epja_2018}. Nevertheless, a systematic analysis of the pair-production process beyond the spatially-uniform-field approximation (we will also call it the dipole approximation) still has not been conducted.

\indent In fact, the previously used dipole approximation (DA) can hardly be justified due to the presence of the fast pulse. The usual ansatz approximating the monochromatic external electric field by a uniform background is justified by the requirement that the laser wavelength $\lambda$ be much larger than the characteristic length scale of the pair-production process $\ell = 2mc^2/|eE|$. This is equivalent to the condition $\gamma \ll 1$ which is not satisfied in the case of a weak and fast pulse since $\gamma_\varepsilon \gg 1$. One may expect that in the presence of both the strong and the weak components, the relevant parameter is the ``combined'' Keldysh parameter $\gamma_\text{c} = m c \omega/|eE|$, but as was demonstrated in a number of studies (see, e.g., Refs.~\cite{schutzhold_prl_2008, orthaber_plb_2011, hebenstreit_plb_2014, linder_prd_2015}), the efficient dynamical assistance is likely to occur only when $\gamma_\text{c} \gtrsim1$. This suggests that the spatial variations of the weak fast pulse should be taken into account, which is the main goal of the present investigation.

\indent In this study we consider a combination of a uniform time-dependent strong field and a standing wave containing rapid oscillations in space and time. Both pulses have a finite duration. We examine the key aspects of the dynamically assisted Schwinger mechanism both within the dipole approximation and beyond it (bDA). According to the results of Ref.~\cite{linder_prd_2015}, the particle yield is exponentially suppressed, and the corresponding exponent does not change when one goes beyond the uniform-external-field approximation. Nevertheless, in this study we carry out numerical calculations which provide the exact values of the number density of particles created, while the worldline instanton approach employed in Ref.~\cite{linder_prd_2015} allows one only to estimate the total particle yield. Besides, we take into account the temporal dependence of the strong pulse and examine various characteristics of the pair-production process. In particular, we analyze the momentum spectra of particles created and the integrated number density. The corresponding calculations are performed by means of a nonperturbative numerical technique. It turns out that taking into consideration the spatial dependence of the weak pulse uncovers a few significant features in the momentum spectra which do not appear within the dipole approximation. Furthermore, the enhancement due to the dynamical assistance as well as the total particle yield also notably alters.

\indent After completion of the present investigation we noticed the very recent study~\cite{lv_pra_2018}, where it was demonstrated that the spatial dependence of the external field plays a crucial role in the context of the Breit-Wheeler process, where a combination of two fast-varying laser pulses is considered. It was shown that one can hardly approximate the resulting field of two pulses with large $\gamma$ by a spatially uniform background. In Ref.~\cite{aleksandrov_prd_2017_2} this conclusion was drawn regarding a combination of two pulses with $\gamma \sim 1$. In the present study, we demonstrate that the same applies to the case of the dynamically assisted Schwinger effect.

\indent In Sec.~\ref{sec:enhancement} we describe the field configuration to be studied and introduce an approximate enhancement factor which is used to identify the values of the field parameters in the dynamical assistance regime. A similar analysis is carried out beyond the dipole approximation, which reveals a number of new important features. In Secs.~\ref{sec:md_da} and \ref{sec:md_bda}, we turn to the study of the momentum distribution of particles produced within the dipole approximation and beyond it, respectively. In Sec.~\ref{sec:total} we examine the total number of $e^+e^-$ pairs and thus provide the exact quantitative comparison of the two approaches. Finally, in Sec.~\ref{sec:discussion} we discuss the main findings of the study and the future prospects. Relativistic units ($\hbar = 1$, $c = 1$) are employed throughout the paper.
%
\section{Approximate enhancement factor}\label{sec:enhancement}
The external electromagnetic field is described by the following vector potential:
\begin{equation}
A_x (t,z) = F(t) \bigg ( \frac{E}{\Omega} \sin \Omega t + \frac{\varepsilon}{\omega} \sin \omega t  \cos k_z z \bigg ),\quad A_y = A_z = 0,\label{eq:field_gen}
\end{equation}
where $k_z = \omega$ and $F(t)$ is a smooth envelope function ($0 \leq F(t) \leq 1$). This external background can be formed by two pairs of counterpropagating laser pulses with a large number of carrier cycles. The envelope $F(t)$ is chosen in the following form:
\begin{equation}
F(t)=
\begin{cases}
\sin^2 \big [ \frac{1}{2} (\pi N - \Omega |t|) \big ] &\text{if}~~\pi (N-1)/\Omega \leq |t| < \pi N/\Omega,\\
1 &\text{if}~~|t| < \pi (N-1)/\Omega,\\
0 &\text{otherwise}.
\end{cases}\label{eq:envelope}
\end{equation}
Accordingly, the field~(\ref{eq:field_gen}) contains $N$ cycles of the slow laser pulse including switching on and switching off parts of half a cycle each and a flat plateau of $N-1$ cycles. The fast pulse governed by the second term in Eq.~(\ref{eq:field_gen}) contains $(\omega/\Omega) N$ cycles. In what follows, we choose $N = 10$, which guarantees that both pulses contain a large number of cycles, and therefore the external background can be approximated by a sum of two standing waves. Since $\gamma_\text{E} \ll 1$, the strong pulse can be considered as a spatially uniform time-dependent field according to the first term in Eq.~(\ref{eq:field_gen}). We also choose $E = 0.2 E_\text{c}$, $\Omega = 0.02m$, and $\gamma_\varepsilon = 10.0$ and vary $\omega$. This leads to $\gamma_E = 0.1$ and $\gamma_\text{c} = 5\, (\omega/m)$.

Within the dipole approximation, the spatial dependence of the second term in Eq.~(\ref{eq:field_gen}) is neglected by replacing $\cos k_0 z$ with $1$. This dependence can be partially taken into account by averaging the results obtained in the dipole approximation for the amplitude $\varepsilon (z) = \varepsilon \cos k_0 z$ being considered at various positions $z \in [0,~2\pi/\Omega ]$. This approach will be referred to as the local dipole approximation.

The method employed in this study is based on the well-known Furry picture formalism incorporating vacuum instability~\cite{fradkin_gitman_shvartsman}. The external field is assumed to act only within the time interval $t_\text{in} < t < t_\text{out}$. One can demonstrate that the number density of particles produced can be directly extracted from the two specific sets of solutions of the Dirac equation. The {\it in} ({\it out}) solutions are determined by their asymptotic behavior in the region $t < t_\text{in}$ ($t > t_\text{out}$). After propagating a given {\it out} solution backwards in time, we decompose it in terms of the {\it in} solutions and obtain the number density of the particles created 
in the corresponding {\it out} state. Since the external field~(\ref{eq:field_gen}) is periodic  (and monochromatic) in space at each time instant $t$, and it does not depend on $x$ and $y$, a given momentum $p_z$ along the $z$ axis can be changed only by an integer number of $\omega$, while components $p_x$ and $p_y$ are conserved. This allows one to propagate only a discrete set of Fourier components for each one-particle solution. This approach was described in detail in Refs.~\cite{aleksandrov_prd_2016, woellert_prd_2015}. As a result, our computations provide the number density of electrons (positrons) produced per unit volume:
\begin{equation}
n(\boldsymbol{p}) = \frac{(2\pi)^3}{V} \, \frac{\mathrm{d}N_{\boldsymbol{p}, s}}{\mathrm{d}^3\boldsymbol{p}}, \label{eq:density}
\end{equation}
where $\boldsymbol{p}$ is the momentum of the particle and $s=\pm 1$ determines its spin state. Due to the symmetry of the external field, the spectra of particles produced are invariant under the reflection $\boldsymbol{p} \to - \boldsymbol{p}$ and independent of $s$.

The local number density $n(\boldsymbol{p})$ considered at a given point $\boldsymbol{p}$  cannot yield a reliable quantitative measure of the dynamical assistance. In this perspective, the total number of pairs, i.e. the function $n(\boldsymbol{p})$ integrated over $\boldsymbol{p}$, seems to be the most suitable parameter. However, its evaluation becomes very time consuming beyond the dipole approximation. For this reason, we study in more detail the number density integrated over $p_y$ at $p_x=p_z=0$:
\begin{equation}
n_y = \! \int \limits_0^{+\infty} \!\! n(0,p_y,0) \mathrm{d}p_y.\label{eq:n_y}
\end{equation}
The $y$ direction is chosen since the magnetic field, which appears beyond the dipole approximation, is directed along the $y$ axis and does not affect much the $p_y$ distribution computed for the spatially homogeneous configuration. This was confirmed by studying an individual pulse as a uniform background and a standing wave, respectively. It turns out that the momentum spectrum in the transversal direction (either $y$ or $z$) in the former case is more similar to the spectrum along the $y$ direction in the latter case (this fact was also indicated in Ref.~\cite{aleksandrov_prd_2017_2}). Moreover, the integral~(\ref{eq:n_y}) converges faster than the analogous $p_x$ and $p_z$ integrals. We use the parameter $n_y$ as a guide for searching for the domain of the dynamical assistance and then study the effect in more detail by calculating the density $n(\boldsymbol{p})$ and the total number of particles created. We also introduce an approximate enhancement factor $K=n_y(\text{I}+\text{II})/[n_y(\text{I})+n_y(\text{II})]$ where $n_y(\text{I})$ and $n_y(\text{II})$ denote the value of $n_y$ in the case of the individual strong and individual weak pulse, respectively, and $n_y(\text{I+II})$ is associated with the combination of the both pulses.

\begin{figure}[h]
\center{\includegraphics[height=0.4\linewidth]{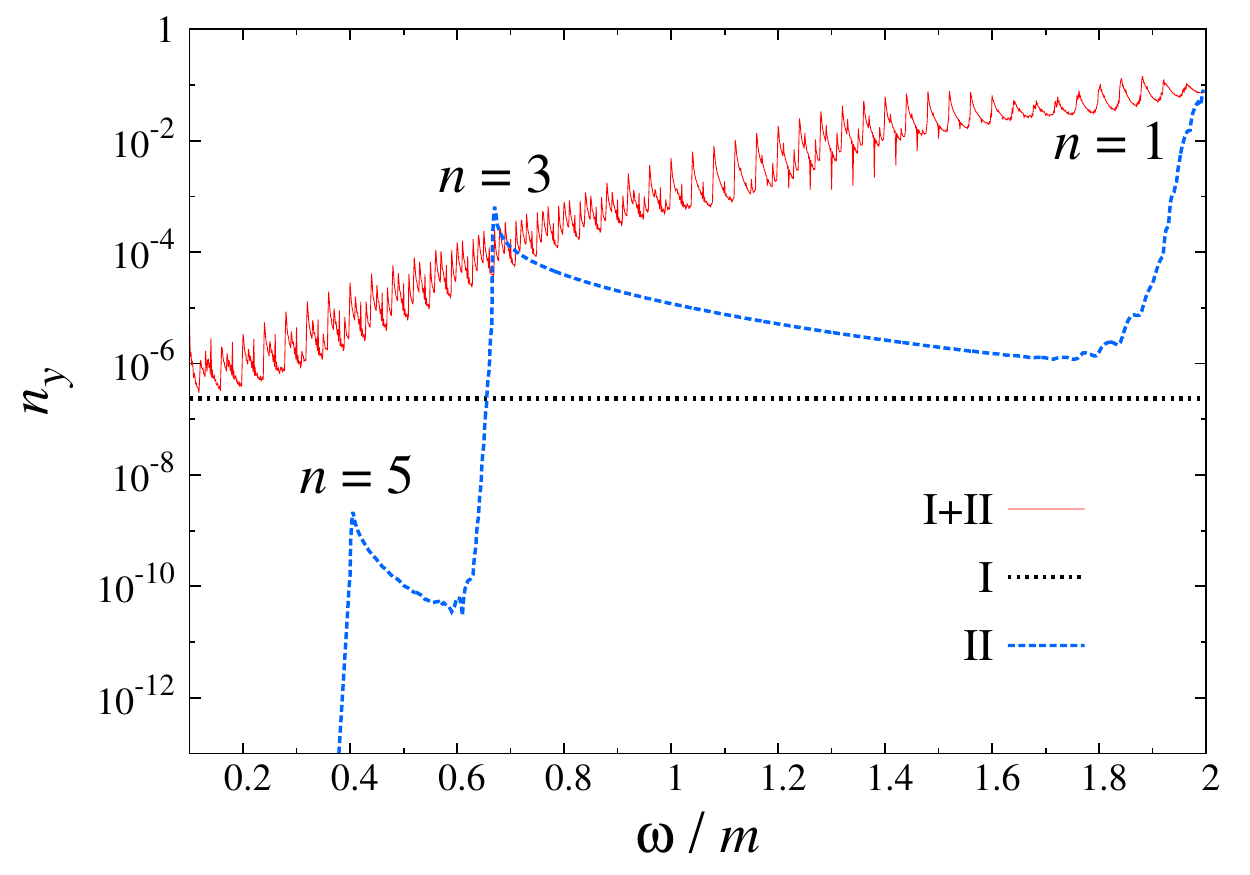}}
\caption{The number density integrated over $p_y$ according to Eq.~(\ref{eq:n_y}) as a function of the fast-pulse frequency $\omega$ in the case of the individual pulses (I and II) and in the presence of both pulses (I+II).}
\label{fig:ny_da}
\end{figure}

Let us first discuss the results obtained within the spatially-uniform-field approximation. In Fig.~\ref{fig:ny_da} we present the values of $n_y$ as a function of the fast-pulse frequency $\omega$ for the case of the individual pulses (I and II) and the combined pulses (I+II). Obviously, the particle yield provided by the strong slow pulse alone (horizontal line) does not depend on $\omega$. On the other hand, the function $n_y (\text{II}) (\omega)$ exhibits a quite nontrivial behavior. Its plot contains a set of large leaps. Each of them corresponds to the appearance of the next $n$-photon channel, and its position can be determined from the condition $2m_* = n \omega$, where $m_*$ is the effective laser-dressed electron mass. In the presence of a weak field ($\gamma_\varepsilon \gg 1$), one has $m_* \approx m$, so the leaps in Fig.~\ref{fig:ny_da} appear at $\omega/m = 2$, $2/3$, $2/5$... The even leaps do not take place here. As was demonstrated in many numerical studies~\cite{mocken_pra_2010, akal_prd_2014, aleksandrov_prd_2017_1, ruf_prl_2009}, the dependence of $n(\boldsymbol{p}=0)$ on $\omega$ has a resonant structure which consists of sharp peaks at $\omega = 2m_*/n$ for odd values of $n$, while the even-$n$ resonances are forbidden. This can be understood if one notes that the angular momentum of the $e^+ e^-$ pair equals zero for $\boldsymbol{p} = 0$, and thus its charge-conjugation parity is $-1$. Since the $C$ parity of the photon is also $-1$, the pair can be generated only by absorbing an odd number of photons. It turns out that this selection rule remains valid even if the transverse component of the particle momentum differs from zero, i.e. $p_y$ and $p_z$ can be arbitrary, provided $p_x = 0$~\cite{mocken_pra_2010, aleksandrov_prd_2017_1, akal_prd_2014}.

\begin{figure}[h]
\center{\includegraphics[height=0.4\linewidth]{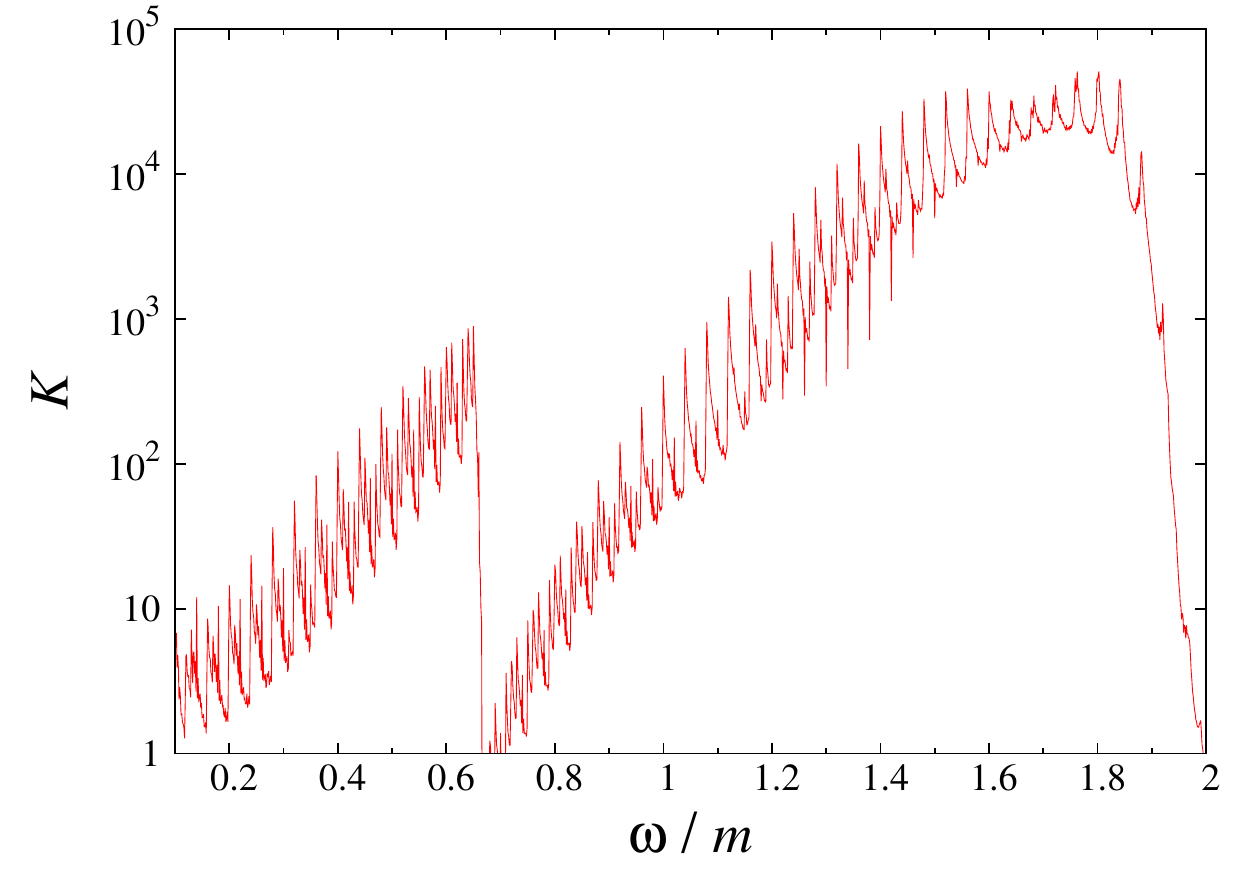}}
\caption{The approximate enhancement factor defined by $K=n_y(\text{I}+\text{II})/[n_y(\text{I})+n_y(\text{II})]$ as a function of the fast-pulse frequency $\omega$.}
\label{fig:K_da}
\end{figure}

When both pulses are present (line ``I+II'' in Fig.~\ref{fig:ny_da}), the pair-production yield becomes substantially larger. In Fig.~\ref{fig:K_da} the approximate enhancement coefficient $K$ is depicted versus $\omega$. One observes that the enhancement can reach several orders of magnitude, but for smaller values of $\omega$, it is also quite small. Furthermore, in order to preserve the nonperturbative character of the pair-production process, one should also make sure that $n_y(\text{I}) \gg n_y(\text{II})$ which holds true only in the region $\omega \lesssim 0.6 m$. This means that the domain of the dynamically assisted Schwinger mechanism is $0.4 m \lesssim \omega \lesssim 0.6 m$.

A special emphasis should be placed on the fact that the more physical characteristic of the pair-production process is the total number of pairs, unlike the rough estimate $n_y$. One should at least verify the findings of such an analysis by the complete calculations of the particle yield. This is especially important for the quantitative comparison of various field configurations and various computational approaches. Besides, the oscillatory behavior of $n_y(\text{I+II})$ (and accordingly $K$) proves to be a nonphysical artifact which does not show up in the total number of particles created. In Sec.~\ref{sec:total} we will address these points in more detail.

\begin{figure}[h]
\center{\includegraphics[height=0.4\linewidth]{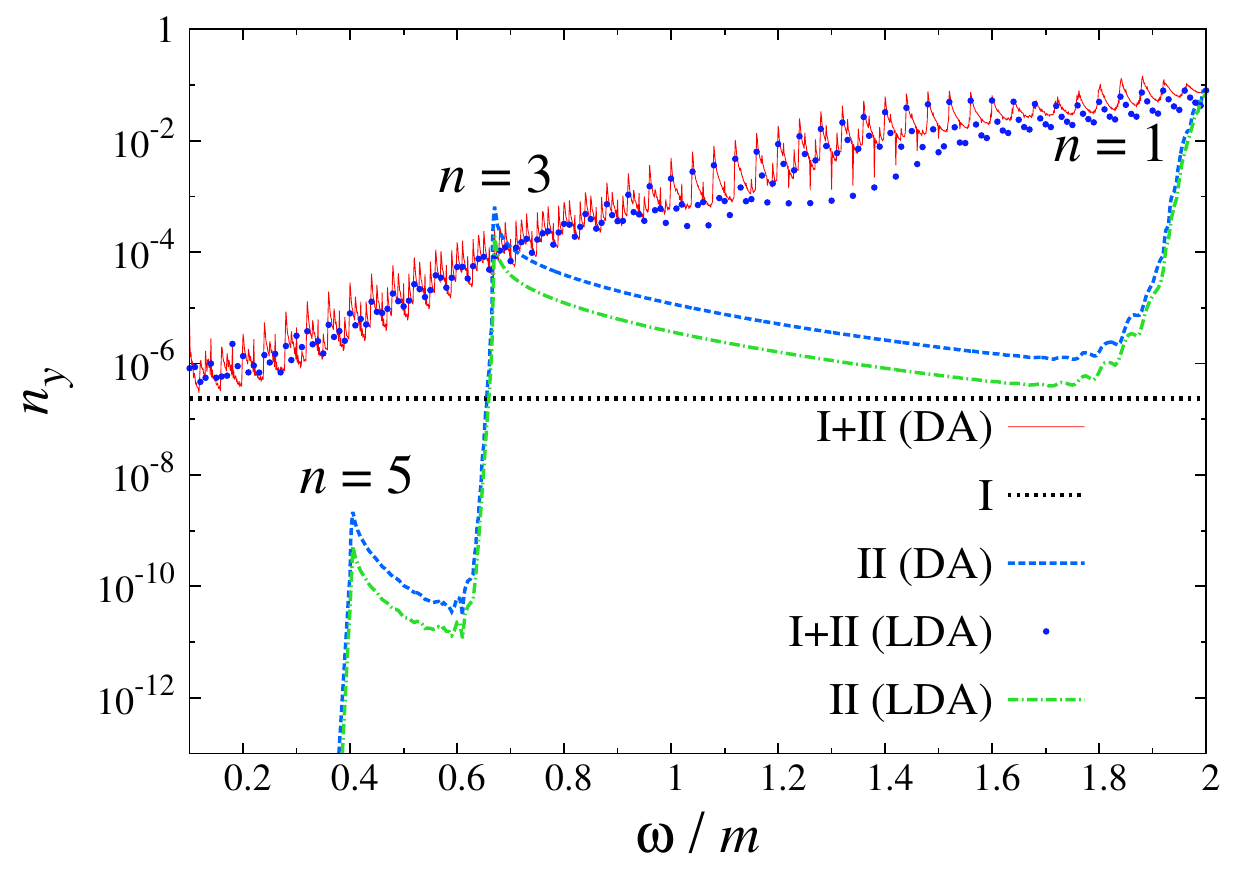}}
\caption{The values of $n_y$ as a function of $\omega$ calculated within the dipole approximation (DA) and the local dipole approximation (LDA) for the three configurations: I, II, and I+II.}
\label{fig:ny_da_lda}
\end{figure}

In Fig.~\ref{fig:ny_da_lda} we present the results obtained within the local dipole approximation. Although they quantitatively differ from the dipole-approximation results for the case of the second pulse alone (II), the qualitative behavior as well as the results for the combined pulses remain almost the same. The analysis of the momentum distribution of particles produced also brings us to the conclusion that the local dipole approximation does not provide any significant findings besides those established in the usual dipole approximation.

\begin{figure}[h]
\center{\includegraphics[height=0.4\linewidth]{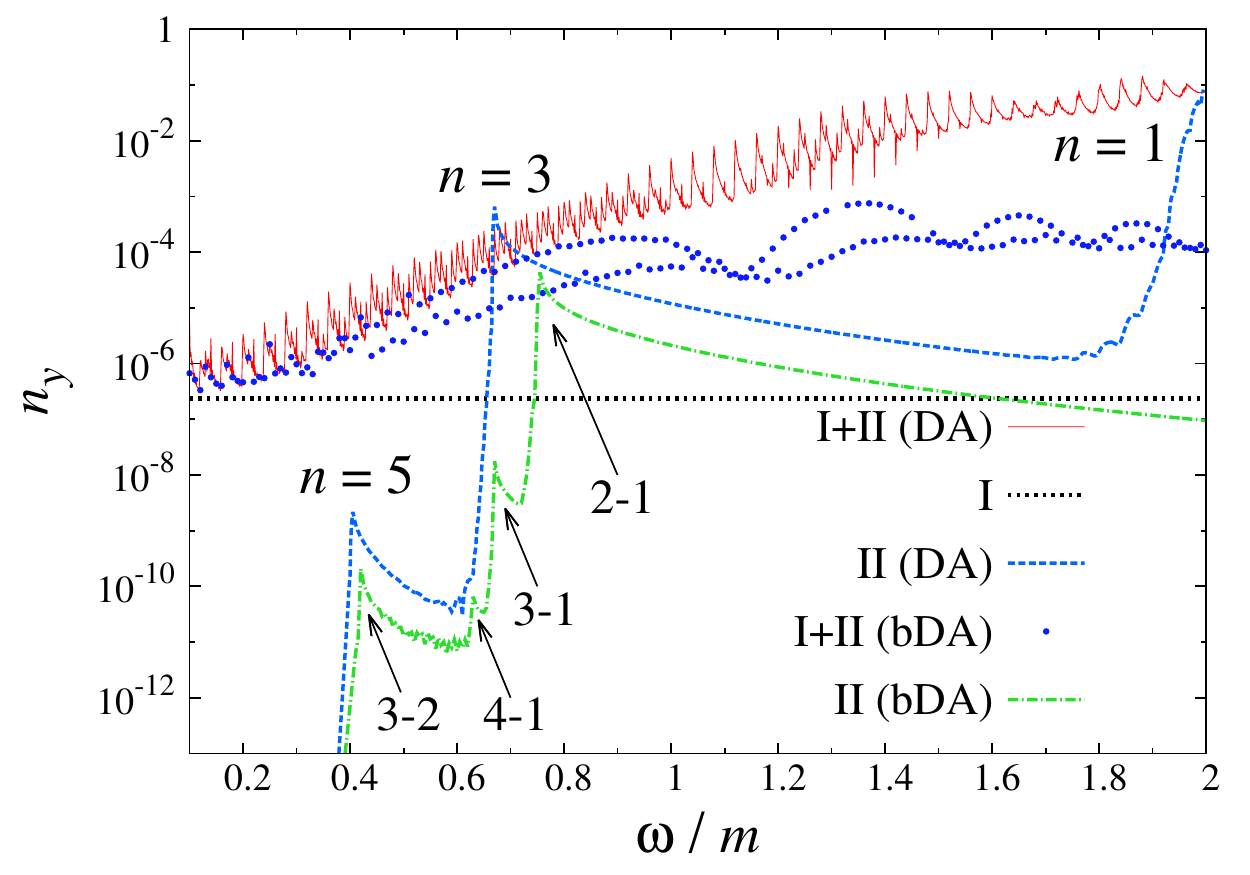}}
\caption{The values of $n_y$ as a function of $\omega$ calculated within the dipole approximation (DA) and beyond it (bDA) for the three configurations: I, II, and I+II.}
\label{fig:ny_da_swa}
\end{figure}

In Fig.~\ref{fig:ny_da_swa} we display the dependences calculated beyond the dipole approximation, i.e. using the expression~(\ref{eq:field_gen}). First, we observe that the dipole approximation considerably overestimates the particle yield, especially in the large-$\omega$ region. It is no surprise since the Keldysh parameter $\gamma_\text{c}$ increases with increasing $\omega$ while the dipole approximation appears to be better justified for smaller $\gamma_\text{c}$. Second, one observes a different multiphoton structure in the case of the individual weak pulse (II). Since the photons now possess not only energy, but also momentum along the $z$ axis (the projection equals $+\omega$ or $-\omega$), the ``resonance'' condition has a different form. Let $q$ and $p$ be the initial and final $4$-momenta of a certain electronic state, respectively. The conservation laws read
\begin{equation}
p = q + n_+ k_+ + n_- k_-, \label{eq:cons_law}
\end{equation}
where $k_{\pm} = (\omega,~0,~0,~\pm \omega )^\text{t}$ and $n_\pm$ are integer numbers.
Taking into account $p_x=p_z=0$ and the relations $p^2 = q^2 = m^2$ (again, the effective mass in a weak field approximately equals the electron mass), one obtains
\begin{equation}
2n_+ n_- \omega = p^0 (n_+ + n_-), \label{eq:res_appr_factor}
\end{equation}
which means that the particle yield should considerably increase with increasing $\omega$ at the points $\omega / m = (n_+ + n_-)/(2n_+n_-)$. One can assume here that $n_+ \geq n_-$. The relation derived now allows one to explain the structure of the graph II (bDA) depicted in Fig.~\ref{fig:ny_da_swa}. The numbers in the graph denote the corresponding values of $n_+$ and $n_-$. A quite similar analysis was performed in Ref.~\cite{ruf_prl_2009} in order to explain the positions of the multiphoton resonances in the scenario involving two counterpropagating high-intensity laser pulses. In Fig.~\ref{fig:ny_da_swa} one notices that beyond the dipole approximation in the vicinity $\omega \approx 2m$ no resonances occur. Another distinctive feature of the $n_y$ dependence consists in the presence of the $3$--$1$ (or $1$--$3$) resonance which corresponds to an even total number of photons absorbed. This demonstrates that the previously discussed selection rule can be violated beyond the dipole approximation.

We observe that the different dynamics taking place beyond the spatially-uniform-field approximation leads to the substantially different patterns (this aspect will also be emphasized in the following two sections). Besides, the more accurate results indicate that the enhancement due to the dynamical assistance is, in fact, weaker. The latter point will also be discussed in Sec.~\ref{sec:total}. In the next two sections, we study the momentum distribution of particles created for the specific choice of $\omega$ (and accordingly $\gamma_\text{c}$).

\section{Momentum distribution within the dipole approximation}\label{sec:md_da}
In this section, we examine the momentum spectra of particles produced within the spatially-uniform-field approximation. The major part of the results is presented for $\omega = 0.5 m$.
\subsection{Transversal direction}\label{subsec:md_da_trans}
\begin{figure}[h]
\center{\includegraphics[height=0.23\linewidth]{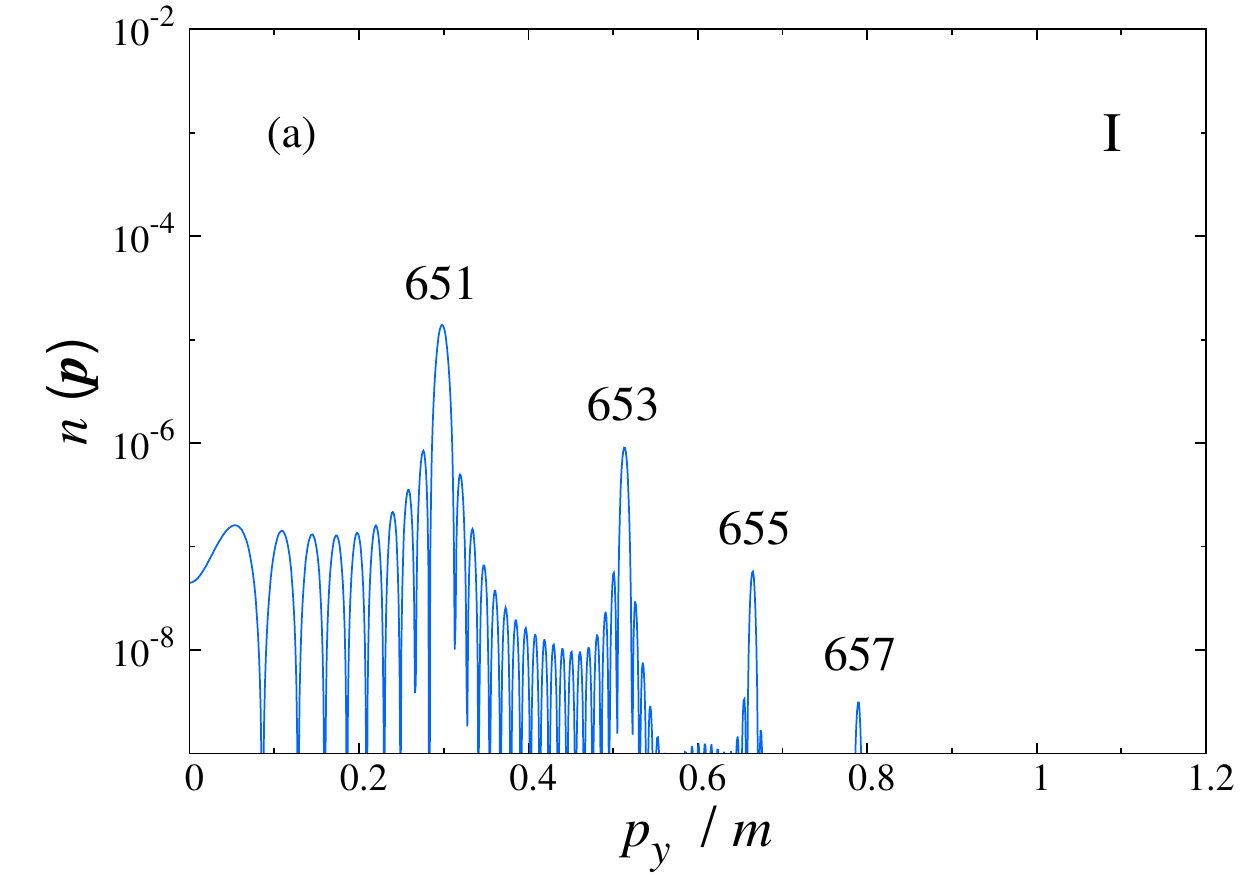}~~\includegraphics[height=0.23\linewidth]{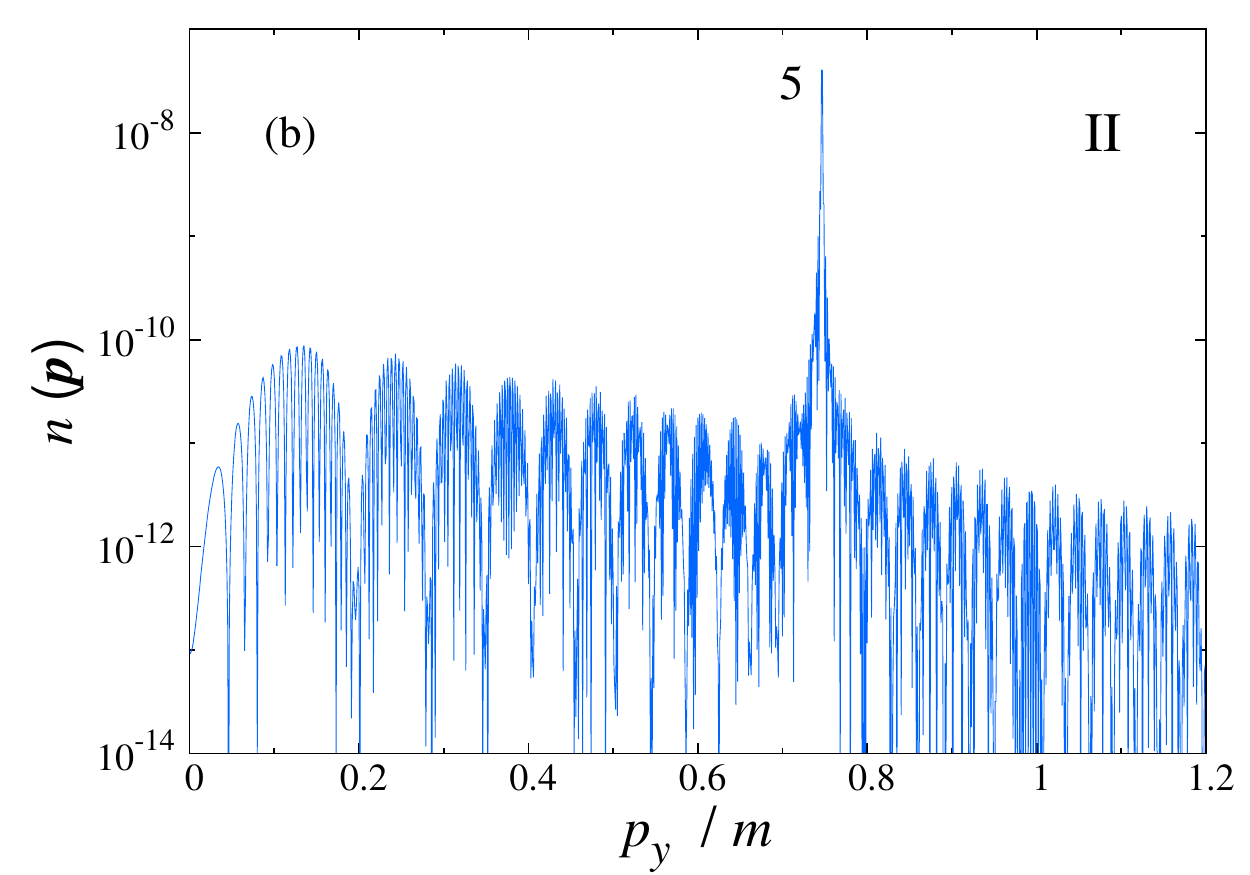}~~\includegraphics[height=0.23\linewidth]{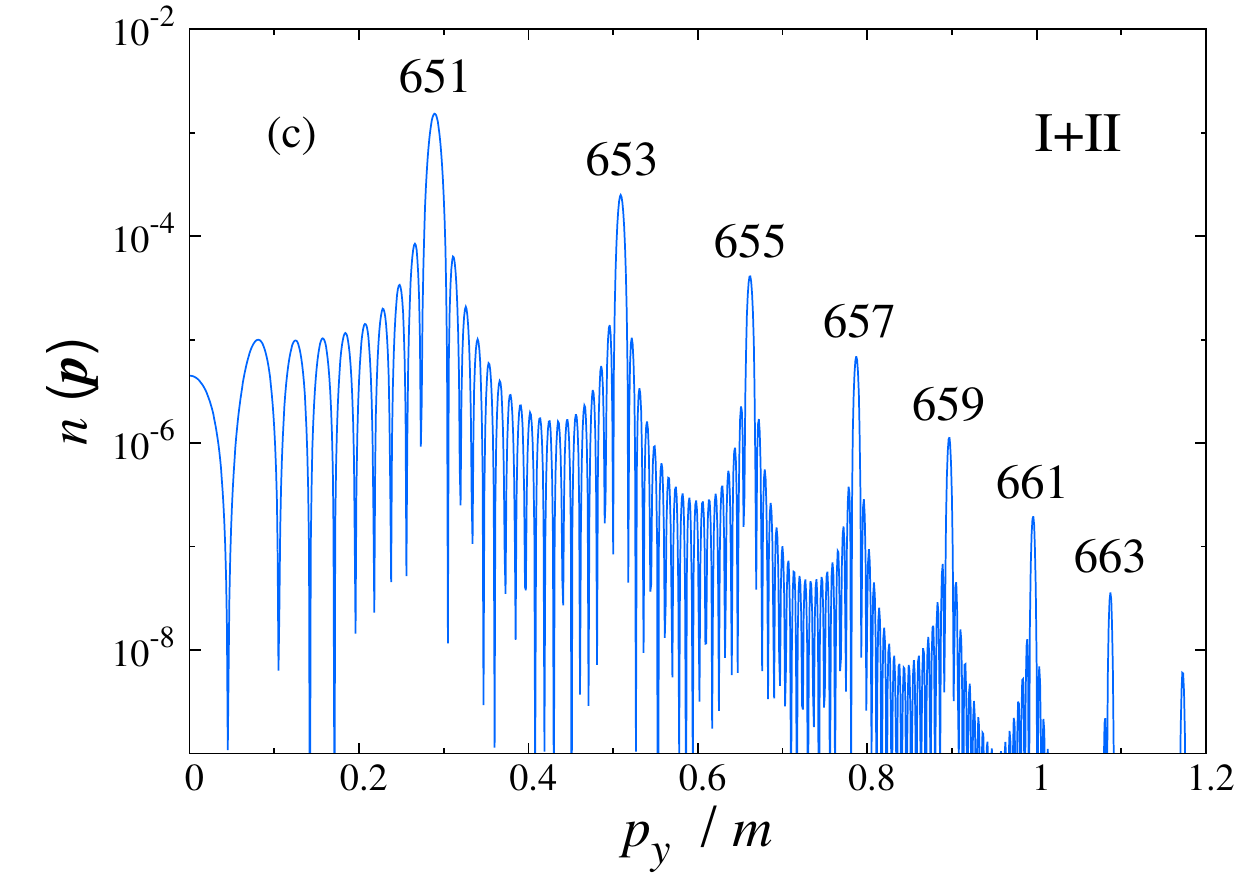}}
\caption{The momentum distribution of particles created as a function of their transversal momentum $p_y$ ($p_x = p_z = 0$) for the three external field configurations (I, II, and I+II) and $\omega = 0.5 m$.}
\label{fig:md_da_trans}
\end{figure}

As was pointed out above, within the dipole approximation, all of the directions in the $y-z$ plane, i.e. perpendicular to the electric field, are equivalent. Without loss of generality, we set $p_z = p_x = 0$ and vary $p_y$. In Fig.~\ref{fig:md_da_trans} we present the momentum distribution of particles created as a function of $p_y$ for the three configurations: I, II, and I+II. The so-called shell structure revealed here was accounted for in Refs.~\cite{otto_plb_2015, otto_prd_2015}. The peaks in Figs.~\ref{fig:md_da_trans}(a) and \ref{fig:md_da_trans}(c) have the positions that satisfy $2\mathcal{E} (0,p_y,0) = n \Omega$ with $\mathcal{E} (\boldsymbol{p})$ being the effective energy in the external field:
\begin{equation}
\mathcal{E} (\boldsymbol{p}) = \frac{1}{2\pi} \int \limits_0^{2\pi} \mathrm{d} x \, \sqrt{m^2+\big [ p_x + \gamma_E^{-1} \sin x + \gamma_\varepsilon^{-1} \sin ( \omega x / \Omega ) \big ]^2 + p_y^2 + p_z^2}, \label{eq:eff_energy}
\end{equation}
where the term with $\gamma_\varepsilon^{-1}$ should be omitted in the case I.
The peaks in Figs.~\ref{fig:md_da_trans}(a) and \ref{fig:md_da_trans}(c) correspond to $n=651$, $653$... In the case II of the weak external background, the effective energy can be estimated as $\mathcal{E} (\boldsymbol{p}) \approx \sqrt{m^2 + \boldsymbol{p}^2}$, and the peak in Fig.~\ref{fig:md_da_trans}(b) is related to the condition $2\sqrt{m^2 + p_y^2} = n \omega$ with $n=5$ ($p_y \approx 0.75 m$). Note that the number of photons is always odd in accordance with the selection rule discussed in the previous section.

The appearance of the fast pulse leads to lifting the momentum distribution evaluated in the case of the slow pulse alone (I). Note that the presence of the weak pulse almost does not affect the expression~(\ref{eq:eff_energy}) since $\gamma_\varepsilon \gg 1$. Accordingly, the lifting effect is not accompanied by any shift of the peaks.

\begin{figure}[h]
\center{\includegraphics[height=0.3\linewidth]{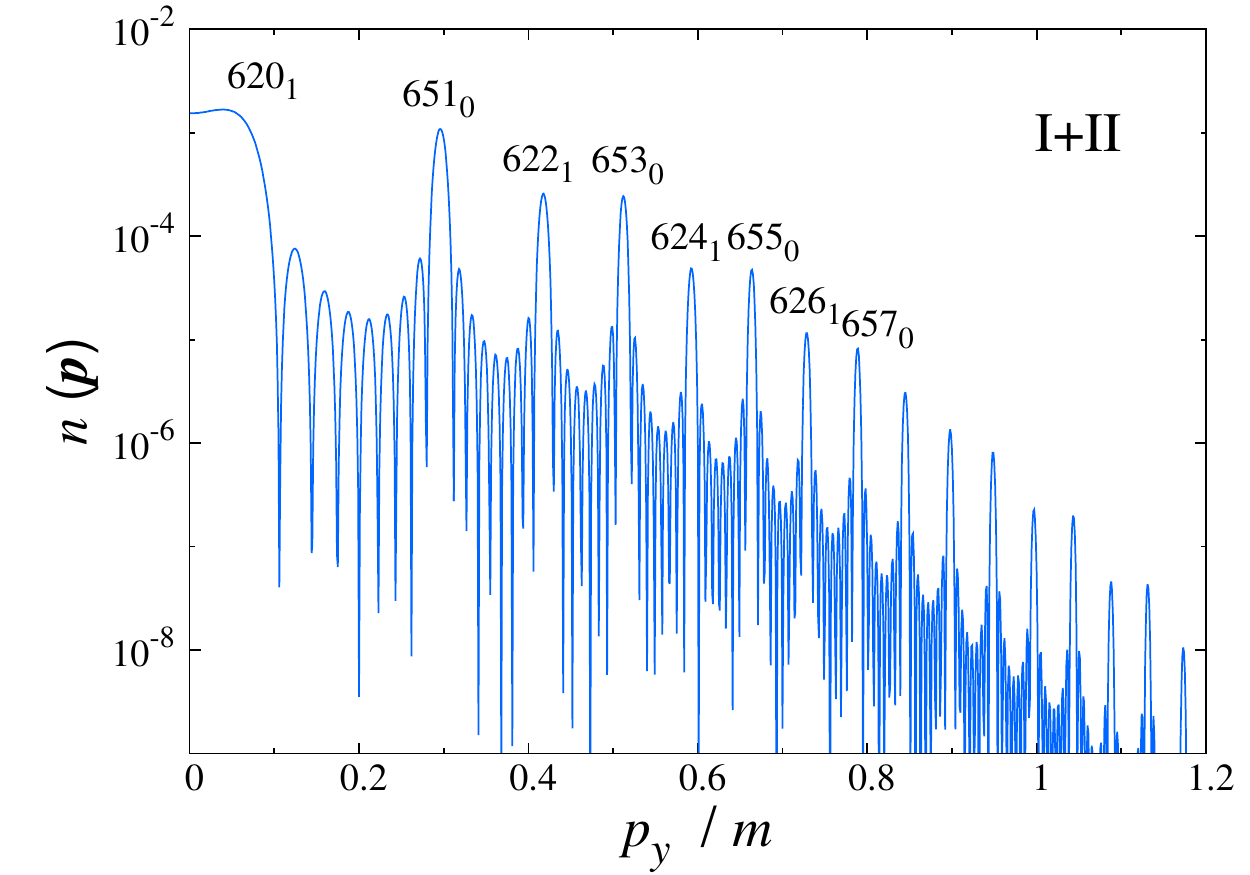}}
\caption{The momentum distribution of particles created as a function of their transversal momentum $p_y$ ($p_x = p_z = 0$) for the combination of the two pulses (I+II) and $\omega = 0.6 m$. The subscripts indicate the number of photons absorbed from the weak pulse.}
\label{fig:md_06_da_trans}
\end{figure}

However, the $e^+ e^-$ pair can be now produced by absorbing $n$ photons of the strong pulse and $\tilde{n}$ photons of the weak one. Supposing that $n$ corresponds to a certain peak in Fig.~\ref{fig:md_da_trans}(a), in the presence of both pulses, the combination of $\tilde{n}$ photons of the weak pulse and $n - (\omega/\Omega)\tilde{n}$ photons of the strong pulse corresponds to the same resonance. Since in our case $\omega/\Omega = 25$, the total number of photons is $n - 24\tilde{n}$, and thus the additional photons of the weak field do not change its parity. This explains why the even resonances do not appear in Fig.~\ref{fig:md_da_trans}(b). Nevertheless, this might as well not be the case. In Fig.~\ref{fig:md_06_da_trans} we display the I+II spectrum for $\omega = 0.6m$.
Since $\omega/\Omega = 30$ is now even, the even peaks now take place, although they do not appear in Fig.~\ref{fig:md_da_trans}(a). The numbers in Fig.~\ref{fig:md_06_da_trans} denote the values of $n$ (large numbers) and $\tilde{n}$ (subscripts). For each resonance, $\tilde{n}$ can be increased by an arbitrary even number $2k$, provided $n$ is decreased by $60k$.
\subsection{Longitudinal direction}\label{subsec:md_da_long}
\begin{figure}[h]
\center{\includegraphics[height=0.23\linewidth]{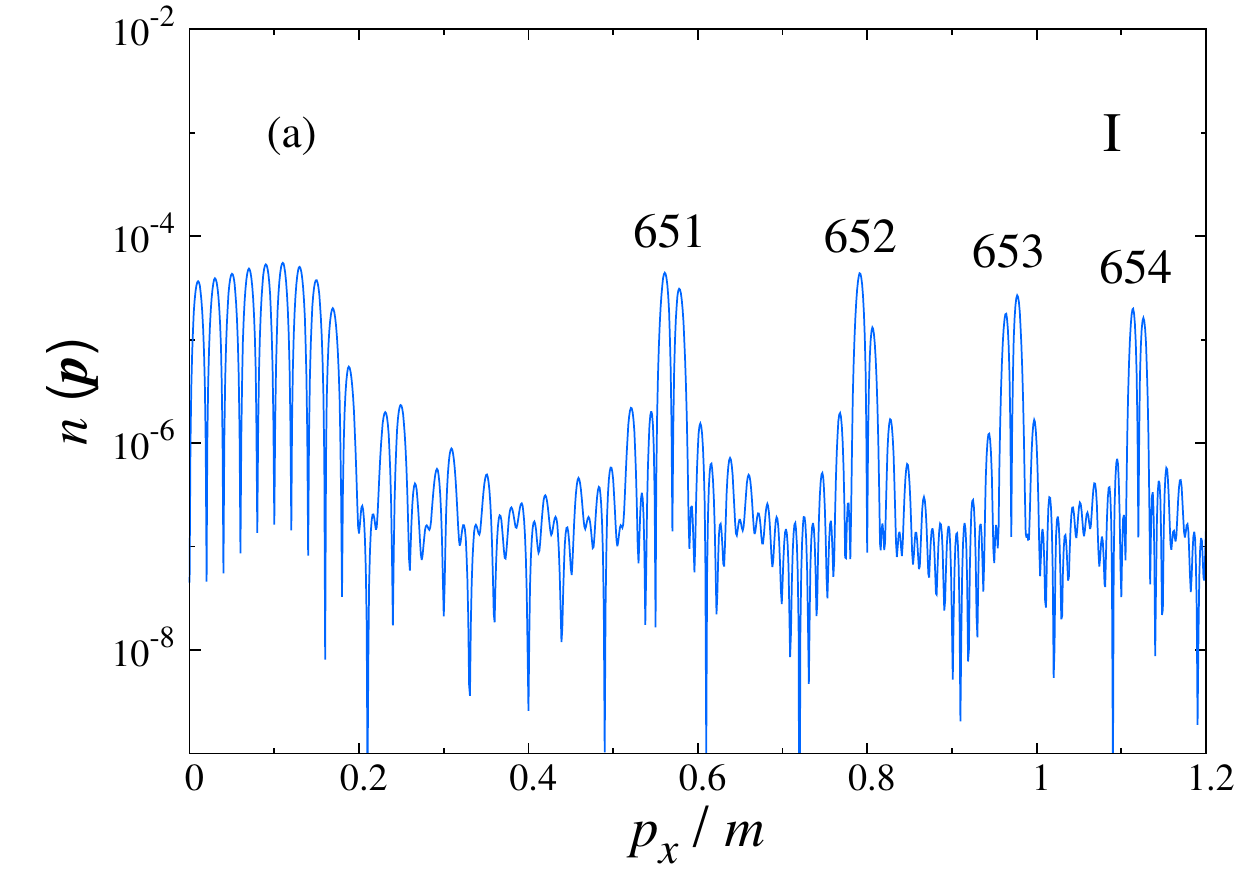}~~\includegraphics[height=0.23\linewidth]{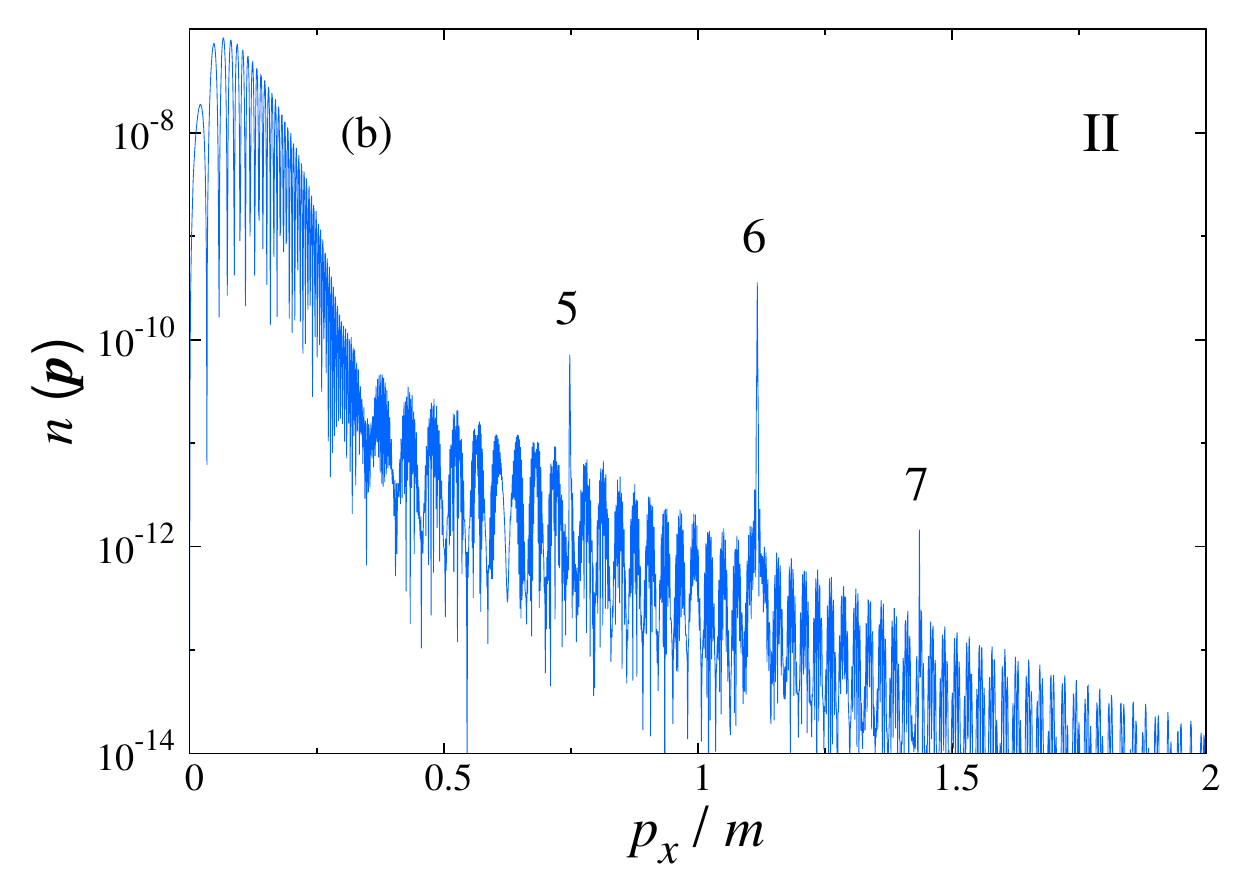}~~\includegraphics[height=0.23\linewidth]{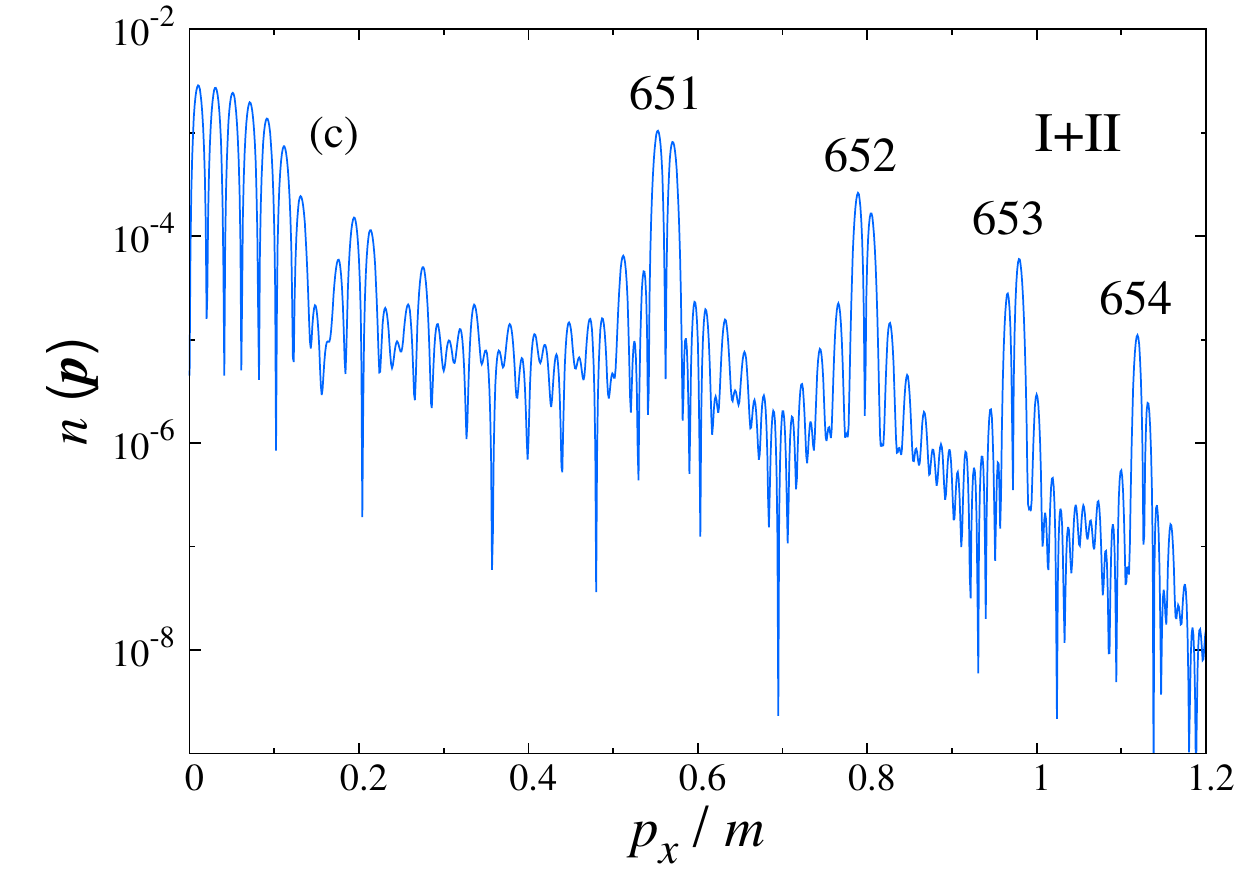}}
\caption{The momentum distribution of particles created as a function of their longitudinal momentum $p_x$ ($p_y = p_z = 0$) for the three external field configurations (I, II, and I+II) and $\omega = 0.5m$.}
\label{fig:md_da_long}
\end{figure}

In Fig.~\ref{fig:md_da_long} we display the spectrum of particles produced with $p_y=p_z=0$ and various values of $p_x$. By means of a similar analysis in terms of resonance conditions, one identifies in Figs.~\ref{fig:md_da_long}(a) and \ref{fig:md_da_long}(c) the peaks with $n = 650$, $651$... The even peaks are now not forbidden. 
In Fig.~\ref{fig:md_da_long}(b) we observe now three sharp peaks which correspond to $n=5$, $6$, and $7$. As $p_x$ tends to $0$ the value of $2\mathcal{E} (p_x,0,0)$ almost reaches $4 \Omega$, which explains the rapid rise of the distribution function. However, at the very point $p_x = 0$ the pair-production probability is again very low. This indicates that the even-$n$ processes are not permitted if the longitudinal momentum vanishes.

Next we will investigate how the patterns discussed above change when one goes beyond the dipole approximation.
\section{Momentum distribution beyond the dipole approximation}\label{sec:md_bda}
The field configuration~(\ref{eq:field_gen}) now consists of both the electric field along the $x$ axis and the magnetic field along the $y$ axis, so the cylindrical symmetry is not present now. In this section we analyze the spectra in the three spatial directions.
\subsection{Magnetic field direction $y$}\label{subsec:md_bda_py}
\begin{figure}[h]
\center{\includegraphics[height=0.3\linewidth]{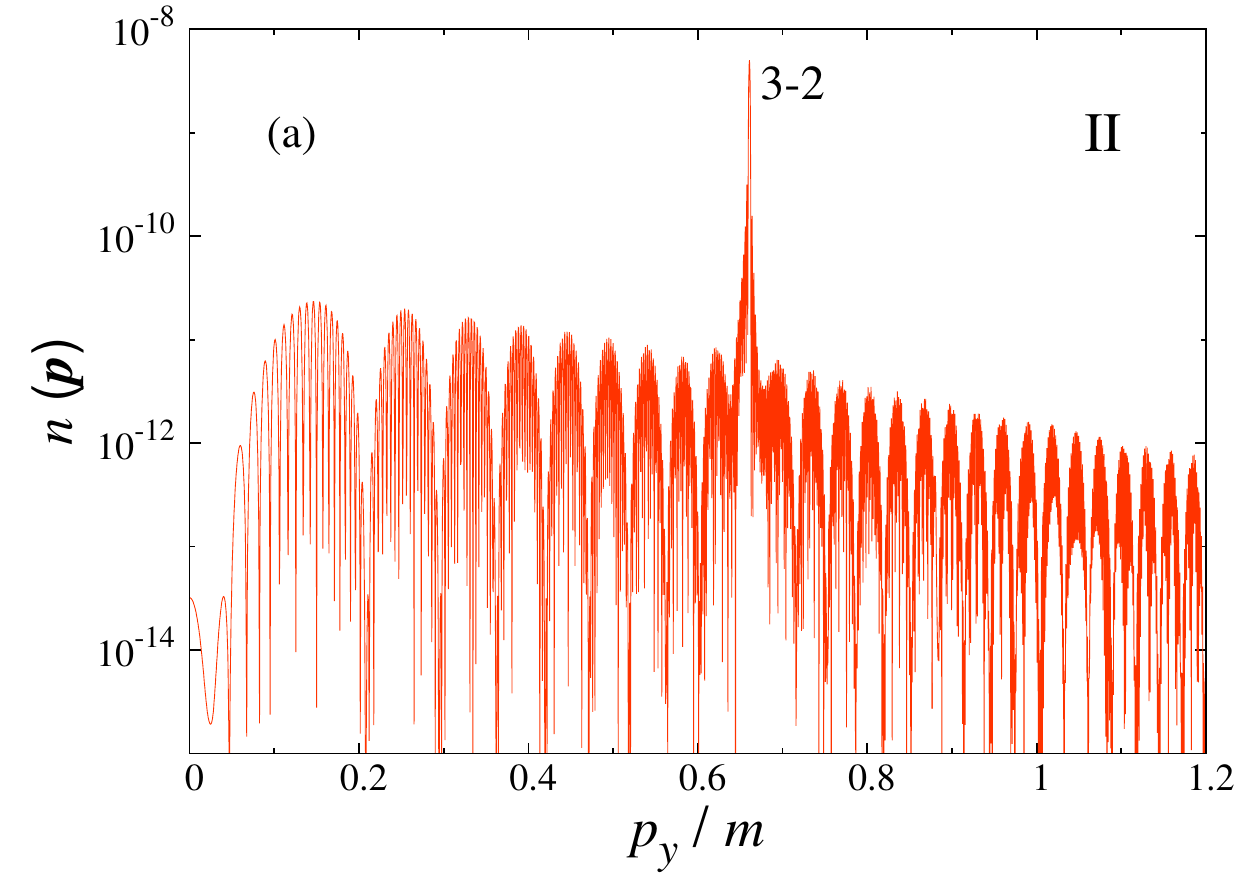}~~\includegraphics[height=0.3\linewidth]{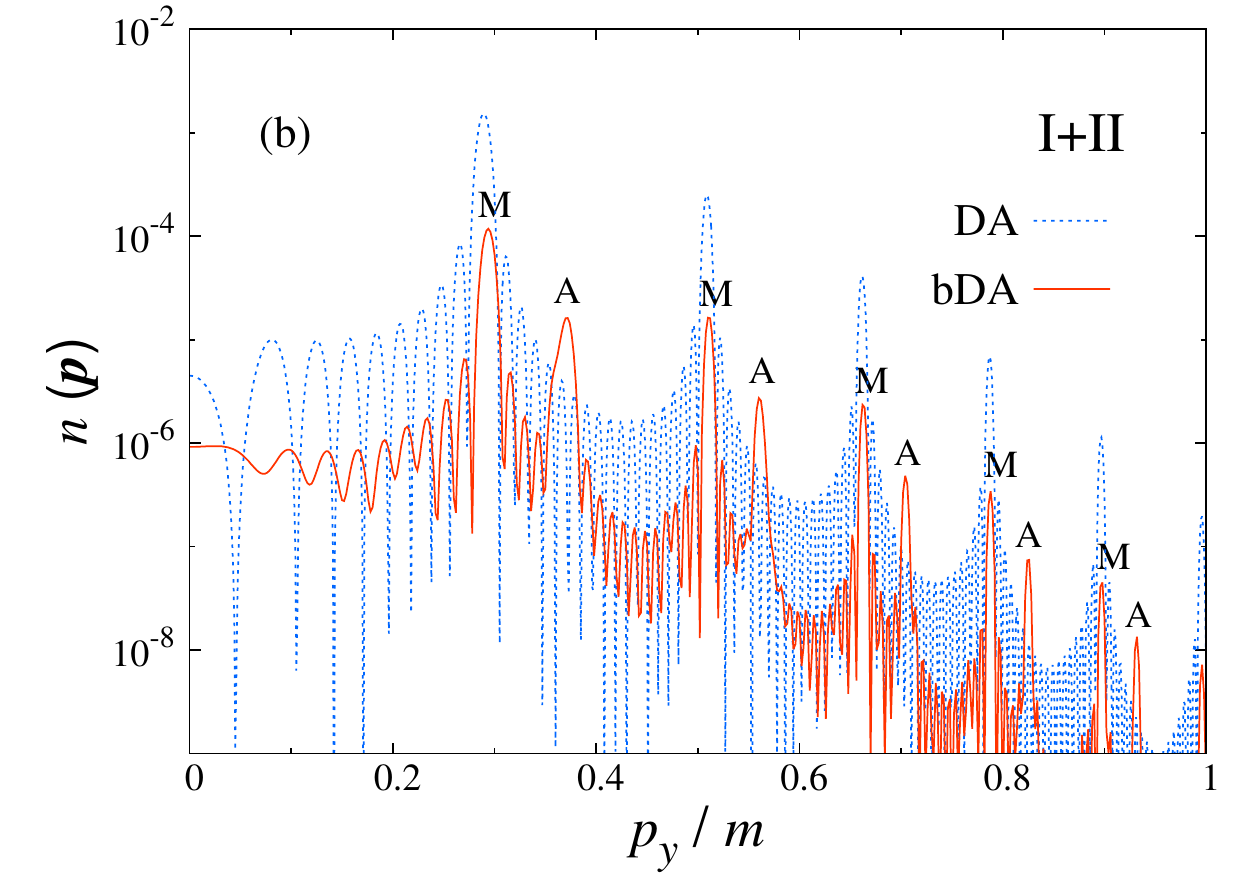}}
\caption{The momentum distribution of particles created as a function of $p_y$ ($p_x = p_z = 0$) for the field configurations II and I+II and $\omega = 0.5m$. The solid lines represent the results obtained beyond the dipole approximation (bDA). In panel (b) the spectrum is compared with the DA results.}
\label{fig:md_bda_py}
\end{figure}

We now set $p_x = p_z = 0$. The $p_y$ spectra contain again a set of pronounced peaks (see Fig.~\ref{fig:md_bda_py}). However, their positions differ from those found in the dipole approximation.
In order to describe this difference in the case of the individual weak pulse [Fig.~\ref{fig:md_bda_py}(a)], we turn again to the conservation law~(\ref{eq:cons_law}). This expression can now be used to determine the resonance position $p_y$ for given $n_+$, $n_-$, and $\omega$. Taking into account $p^2 = q^2 = m^2$, one obtains
\begin{equation}
p_y/m = \sqrt{\frac{(2 n_+ n_-)^2}{(n_+ + n_-)^2}\, \bigg (\frac{\omega}{m}\bigg )^2 - 1}. \label{eq:res_bda_py}
\end{equation}
This expression predicts a resonant peak at $p_y \approx 0.66 m$ (resonance $3$--$2$ or $2$--$3$) which is clearly seen in Fig.~\ref{fig:md_bda_py}(a). The other resonances are considerably suppressed as they appear in higher orders of perturbation theory. The resonance $2$--$2$ would correspond to $p_y = 0$, but it does not show up in Fig.~\ref{fig:md_bda_py}(a) since it has an even sum $n_+ + n_-$. The analysis of the momentum distributions beyond the dipole approximation reveals that the processes with even photon numbers are suppressed only in the case of the $p_y$ spectra.

In the presence of the two pulses, the spectrum possesses a more complicated structure. Besides the peaks predicted within the dipole approximation, there exist also additional peaks in between. They can be accounted for by means of the conservation laws, which in this case take the following form:
\begin{equation}
p = q + n_+ k_+ + n_- k_- + n k_0, \label{eq:cons_law2}
\end{equation}
where $k_0 = (\Omega,~0,~0,~0)^\text{t}$ is the $4$-momentum of the strong pulse photon. Then we set $p_x = p_z = 0$ and use the relations $p^0 = \mathcal{E} (\boldsymbol{p})$ and $q^0 = \mathcal{E} (\boldsymbol{q})$. The resonance condition reads:
\begin{equation}
\mathcal{E} (0, p_y, 0) + \mathcal{E} (0, p_y, (n_- - n_+)\omega) = (n_+ + n_-) \omega + n \Omega. \label{eq:res_bda_two}
\end{equation}
In order to evaluate the effective energy $\mathcal{E} (\boldsymbol{p})$, we employ again the expression~(\ref{eq:eff_energy}) even though we go beyond the dipole approximation. The reason for this is that the second pulse contribution (the term with $\gamma_\varepsilon^{-1}$) is always very small, so it does not need to be modified. Using Eqs.~(\ref{eq:eff_energy}) and (\ref{eq:res_bda_two}), we identify the resonant peaks in Fig.~\ref{fig:md_bda_py}(b). It turns out that the main peaks, which can also be found in Figs.~\ref{fig:md_da_trans}(a) and \ref{fig:md_da_trans}(c), correspond to the processes with $n_+ = n_-$. For each value of $n_+ = n_-$, there is the same series of main peaks being enumerated by $n$ (see Table~\ref{table:1}). The additional peaks in Fig.~\ref{fig:md_bda_py}(b) emerge as the resonances with $n_+ \neq n_-$. Note that Eq.~(\ref{eq:res_bda_two}) is symmetric with respect to the interchange $n_+ \leftrightarrow n_-$, so we assume that $n_+ \geq n_-$.
\begin{table}[h]
\centering
\setlength{\tabcolsep}{0.5em}
\begin{tabular}{l|c|c}
\hline
\hline
Series                            & $n_+$ -- $n_-$   & $n$   \\ \hline
\multirow{3}{*}{Main peaks (M)}       & $0$ -- $0$ & $651$, $653$, ... \\
                                  & $1$ -- $1$ & $601$, $603$, ... \\
                                  & ... & ... \\ \hline
\multirow{6}{*}{Additional peaks (A)} & $1$ -- $0$ & $628$, $630$, ... \\
                                  & $2$ -- $1$ & $578$, $580$, ... \\
                                  & ... & ... \\ \cline{2-3} 
                                  & $2$ -- $0$ & $607$, $609$ ... \\
                                  & $3$ -- $1$ & $557$, $559$, ... \\
                                  & ... & ... \\ \hline \hline
\end{tabular}
\caption{The series of the resonant peaks in Fig.~\ref{fig:md_bda_py}(b). Each of the M series predicts the main peaks already found within the dipole approximation while all of the A series reproduce the additional ones.}
\label{table:1}
\end{table}
The resonance condition~(\ref{eq:res_bda_two}) formally allows the integers $n_+$ and $n_-$ to also be negative. This, however, in turn, leads to greater values of $n$, and thus such processes are strongly suppressed in comparison to those displayed in Table~\ref{table:1} and thus are not indicated here. Note that the spectrum contains only the peaks with an odd sum $n+n_++n_-$. The resonances located by means of Eq.~(\ref{eq:res_bda_two}) and those found numerically coincide at least with $1.5 \%$ accuracy.

If the additional peaks appear already in the DA spectrum, e.g. for $\omega = 0.6 m$ (see Fig.~\ref{fig:md_06_da_trans}), the results obtained beyond the DA reproduce the same resonant structure. If the DA distribution contains only odd peaks, the number of resonances doubles beyond this approximation. Although the resonant structure appears mainly owing to the presence of the high-intensity slow field, the modified dynamics of the weak pulse beyond the dipole approximation gives rise to the additional signatures in the momentum spectrum. The weak fast pulse now not only lifts the momentum distribution but also changes its overall structure assisting the pair production process in the strong field.

\subsection{Propagation direction $z$}\label{subsec:md_bda_pz}
\begin{figure}[h]
\center{\includegraphics[height=0.3\linewidth]{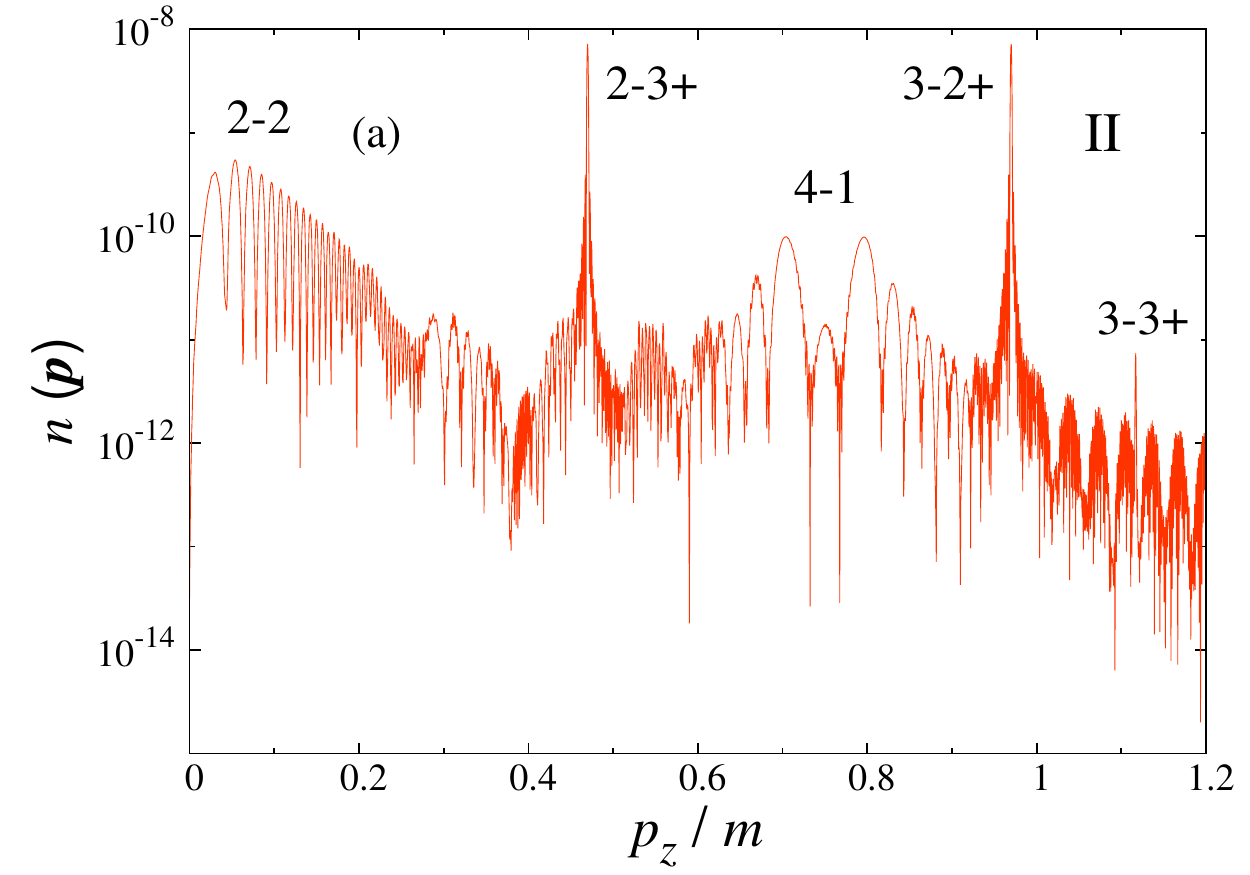}~~\includegraphics[height=0.3\linewidth]{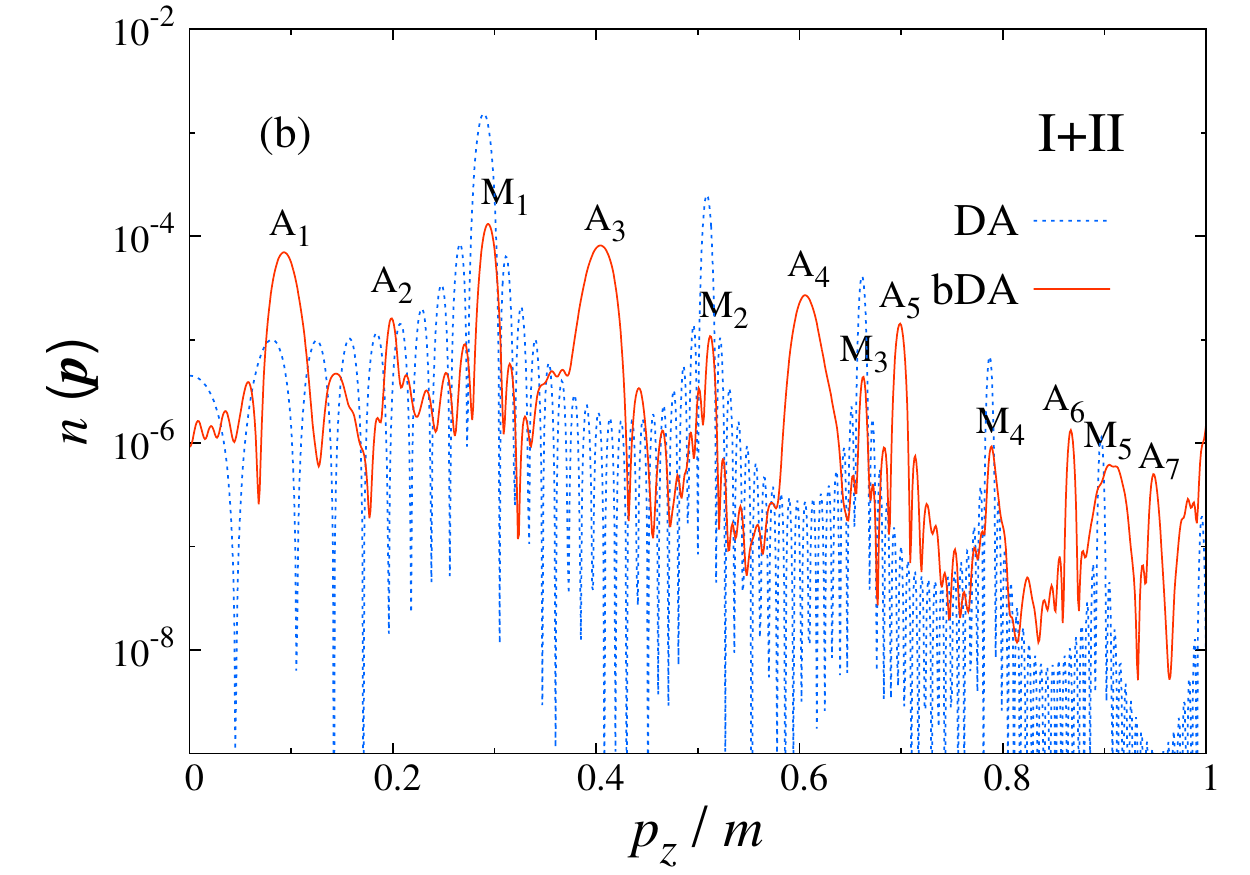}}
\caption{The momentum distribution of particles created as a function of $p_z$ ($p_x = p_y = 0$) for the field configurations II and I+II ($\omega = 0.5m$).}
\label{fig:md_bda_pz}
\end{figure}

When only the weak pulse is present, the spectrum contains peaks which can be located using Eq.~(\ref{eq:cons_law}) [see Fig.~\ref{fig:md_bda_pz}(a)]. However, the resonant values of $p_z$ are now not described by the right-hand side of Eq.~(\ref{eq:res_bda_py}) since the $p_z$ component of the particle momentum can change due to the absorption of photons. Setting $p_x = p_y = 0$ and using Eq.~(\ref{eq:cons_law}), one obtains:
\begin{equation}
2n_+ n_- \omega = p^0 (n_+ + n_-) - p_z (n_+ - n_-), \label{eq:res_bda_pz_1}
\end{equation}
where $p^0 = \sqrt{m^2 + p_z^2}$. This leads to
\begin{equation}
p_z = \frac{n_+ - n_-}{2} \, \omega \pm \frac{n_+ + n_-}{2} \sqrt{\omega^2 - \frac{m^2}{n_+ n_-}}, \label{eq:res_bda_pz_2}
\end{equation}
where $n_+$ and $n_-$ are positive. This expression allows one to identify the resonances in Fig.~\ref{fig:md_bda_pz}(a).
Whereas in the dipole approximation one observes only one peak at $p_y = 0.75m$, beyond the DA, the spectrum becomes more complicated. The two high sharp peaks are associated with the $2$--$3+$ and $3$--$2+$ transitions, where, besides $n_+$ and $n_-$, we indicate the sign in Eq.~(\ref{eq:res_bda_pz_2}). Note that both the $4$--$1+$ and $4$--$1-$ resonances correspond to $p_z = 0.75 m$ because the square root in Eq.~(\ref{eq:res_bda_pz_2}) vanishes. Moreover, according to Eq.~(\ref{eq:res_bda_pz_2}), the positions of these ``accidentally'' degenerate resonances are very sensitive with respect to the small changes of the fast-pulse frequency $\omega$. It turns out that for $\omega = 0.50034 m$ the expression (\ref{eq:res_bda_pz_2}) predicts the $4$--$1+$ and $4$--$1-$ peaks at $p_z = 0.797 m$ and $p_z = 0.704 m$, respectively, which correspond to the peaks in Fig.~\ref{fig:md_bda_pz}(a). On the other hand, the positions of the other resonances change by less than $0.3\%$. Since the external field~(\ref{eq:field_gen}) is, in fact, not monochromatic, it is no accident that the $4$--$1$ peak splits into two. One could also expect that the structure of the momentum distribution in Fig.~\ref{fig:md_bda_pz}(a) is quite unstable in the vicinity of $p_z = 0.75 m$. Our computations with different envelope functions $F(t)$ confirm this point. The same holds true when one analyzes the peaks $2$--$2+$ and $2$--$2-$ in the vicinity of $p_z = 0$. In contrast to the results obtained in the dipole approximation, the even resonances are now allowed.

\begin{figure}[h]
\center{\includegraphics[height=0.3\linewidth]{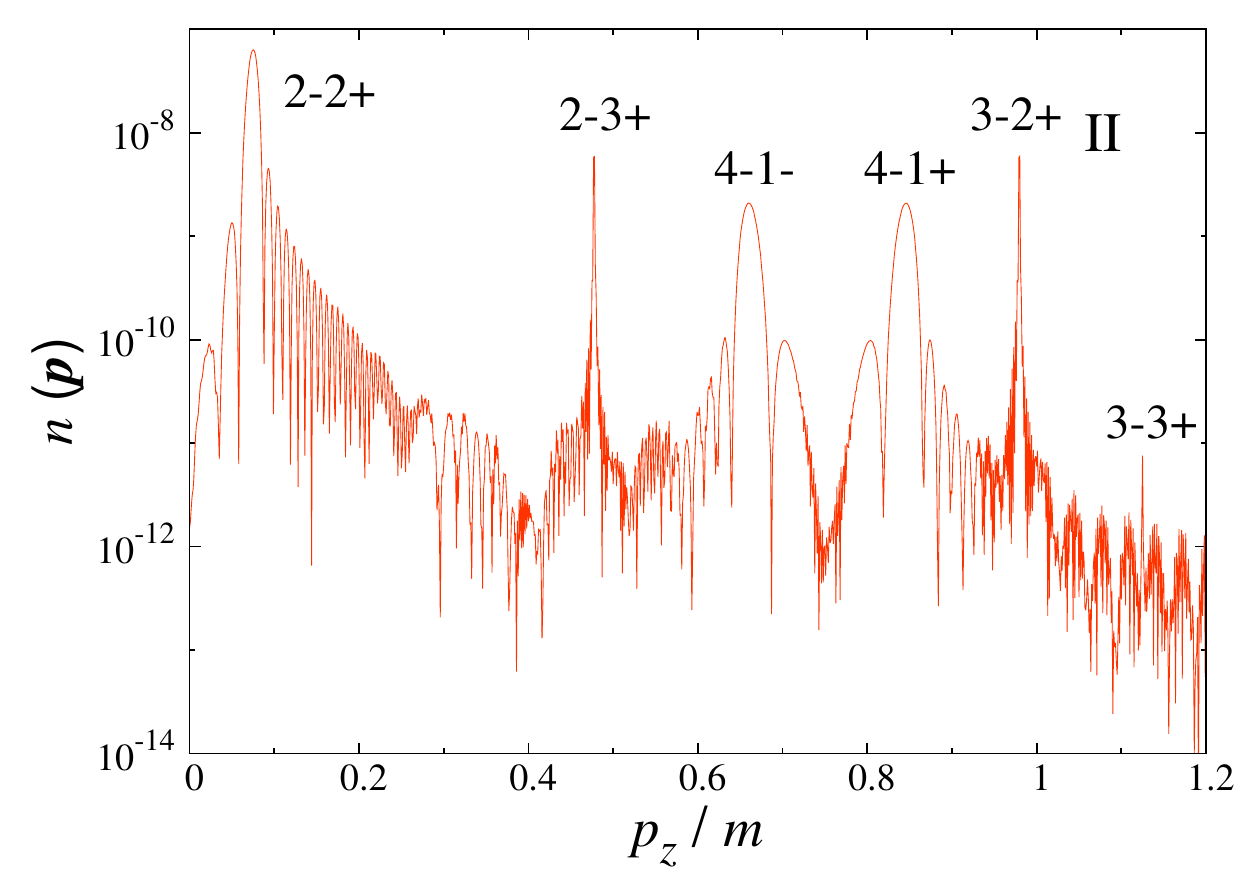}}
\caption{The momentum distribution of particles created as a function of $p_z$ ($p_x = p_y = 0$) for the field configuration II and $\omega = 0.502m$.}
\label{fig:md_bda_pz_0502}
\end{figure}

To further clarify and illustrate the aspects discussed, we present the spectrum for $\omega = 0.502m$ (see Fig.~\ref{fig:md_bda_pz_0502}).
The $2$--$2$ and $4$--$1$ resonances split and form four distinct peaks (the $2$--$2-$ peak has a negative value of $p_z$), while the positions of the peaks $2$--$3+$, $3$--$2+$, and $3$--$3+$ remain almost the same.

In the presence of the two pulses [see Fig.~\ref{fig:md_bda_pz}(b)], the resonant structure can be deciphered as in the previous subsection. Instead of Eq.~(\ref{eq:res_bda_two}), one has now
\begin{equation}
\mathcal{E} (0, 0, p_z) + \mathcal{E} (0, 0, p_z + (n_- - n_+)\omega) = (n_+ + n_-) \omega + n \Omega. \label{eq:res_bda_two_pz}
\end{equation}
This relation does not possess the symmetry $n_+ \leftrightarrow n_-$ and provides now a larger variety of resonances. The peaks in Fig.~\ref{fig:md_bda_pz}(b) are described in Table~\ref{table:2}.
\begin{table}[h]
\centering
\setlength{\tabcolsep}{0.5em}
\begin{tabular}{c|c|c|c|c|c|r}
\hline
\hline
Peak & $n$ & $n_+$ & $n_-$ & $p_z/m$ (D) & $p_z/m$ (E) & \multicolumn{1}{c}{$n (\boldsymbol{p})$} \\ \hline
$\text{M}_1$ & $651$ & $0$ & $0$ & $0.293$ & $0.294$ & $1.3\times 10^{-4}$ \\
$\text{M}_2$ & $653$ & $0$ & $0$ & $0.511$ & $0.512$ & $1.1\times 10^{-5}$ \\
$\text{M}_3$ & $655$ & $0$ & $0$ & $0.663$ & $0.663$ & $4.4\times 10^{-6}$ \\
$\text{M}_4$ & $657$ & $0$ & $0$ & $0.788$ & $0.789$ & $9.3\times 10^{-7}$ \\
$\text{M}_5$ & $659$ & $0$ & $0$ & $0.898$ & $0.905$ & $6.2\times 10^{-7}$ \\
$\text{A}_1$ & $626$ & $1$ & $0$ & $0.096$ & $0.093$ & $7.0\times 10^{-5}$ \\
$\text{A}_2$ & $628$ & $0$ & $1$ & $0.198$ & $0.199$ & $1.6\times 10^{-5}$ \\
$\text{A}_3$ & $626$ & $1$ & $0$ & $0.404$ & $0.404$ & $8.2\times 10^{-5}$ \\
$\text{A}_4$ & $603$ & $2$ & $0$ & $0.607$ & $0.606$ & $2.7\times 10^{-5}$ \\
$\text{A}_5$ & $628$ & $1$ & $0$ & $0.698$ & $0.699$ & $1.4\times 10^{-5}$ \\
$\text{A}_6$ & $630$ & $1$ & $0$ & $0.868$ & $0.867$ & $1.3\times 10^{-6}$ \\
$\text{A}_7$ & $605$ & $2$ & $0$ & $0.946$ & $0.948$ & $5.0\times 10^{-7}$ \\ \hline \hline
\end{tabular}
\caption{The list of the resonant peaks discovered beyond the dipole approximation in the $p_z$ spectrum [Fig.~\ref{fig:md_bda_pz}(b)]. The $p_z$ values derived from~Eq.~(\ref{eq:res_bda_two_pz}) (D) match those found exactly (E).}
\label{table:2}
\end{table}
Each $n$--$n_+$--$n_-$ resonance can also be represented as the $(n-50)$--$(n_++1)$--$(n_-+1)$ resonance similarly to what is shown in Table~\ref{table:1}. Although in Table~\ref{table:2} the total number of photons $n+n_++n_-$ is always odd, the even peaks can also emerge as was found in our calculations for other values of $\omega$.
\subsection{Electric field direction $x$}\label{subsec:md_bda_px}
\begin{figure}[h]
\center{\includegraphics[height=0.3\linewidth]{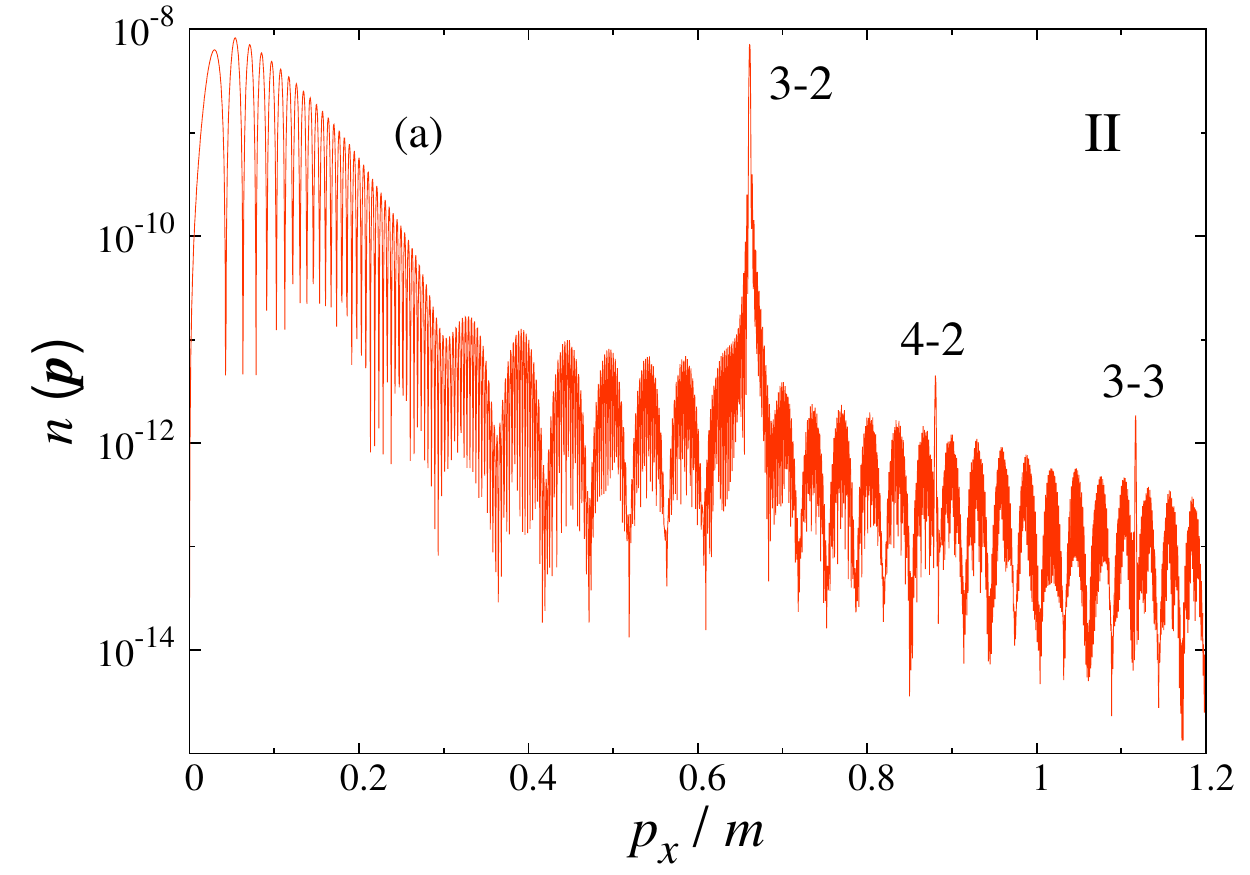}~~\includegraphics[height=0.3\linewidth]{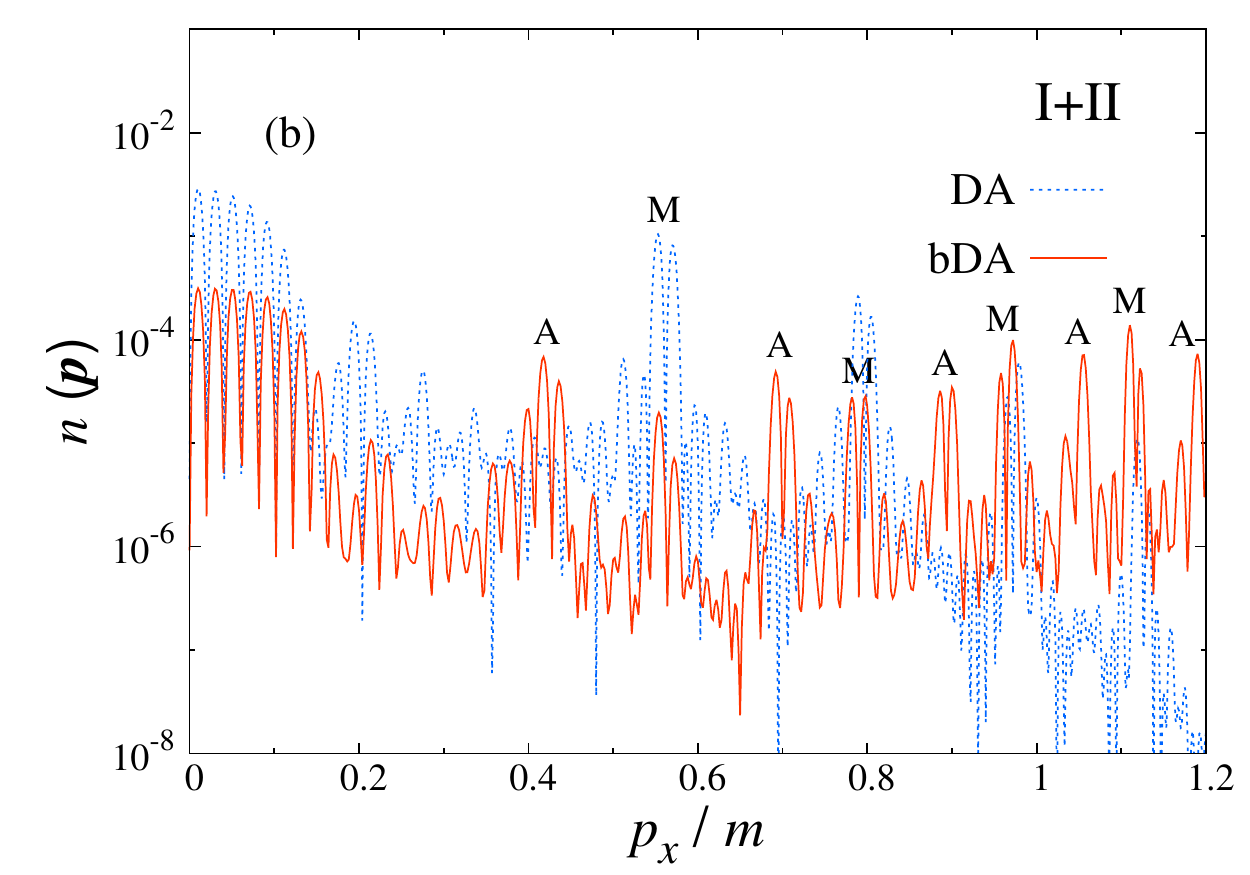}}
\caption{The momentum distribution of particles created as a function of $p_x$ ($p_y = p_z = 0$) for the field configurations II and I+II ($\omega = 0.5m$).}
\label{fig:md_bda_px}
\end{figure}

The momentum distributions for $p_y=p_z=0$ are depicted in Fig.~\ref{fig:md_bda_px}. Their structure can be explained almost in the same way as it was done for the $p_y$ spectrum.
In the case of the fast pulse (II), the only difference is that the even resonances are now not forbidden, so the $2$--$2$ resonance now leads to a dramatic rise of the production probability in the vicinity of $p_x = 0$ [Fig.~\ref{fig:md_bda_px}(a)]. The peaks $4$--$2$ ($2$--$4$) and $3$--$3$ are less pronounced, for they correspond to higher orders of perturbation theory.

The resonant structure in Fig.~\ref{fig:md_bda_px}(b) is notably modified in comparison to the results obtained in the dipole approximation. Apart from the previously found (main) resonances, there are again additional peaks. The resonance condition now reads:
\begin{equation}
\mathcal{E} (p_x, 0, 0) + \mathcal{E} (p_x, 0, (n_- - n_+)\omega) = (n_+ + n_-) \omega + n \Omega. \label{eq:res_bda_two_px}
\end{equation}
By inspection of this equation, we find that the additional peaks correspond to the absorption of one fast-pulse photon traveling in either direction and $n=627$, $628$... Alternatively, the resonances can appear as the $2-1$ (or $1-2$) processes with $n=577$, $578$... or in even higher orders in $n_+$ and $n_-$.

Performing the more accurate calculations beyond the uniform-field approximation, we establish that the momentum spectra of particles have in fact a different structure. Nevertheless, the accurate quantitative comparison of the two approaches seems complicated. For instance, in Fig.~\ref{fig:md_bda_px}(b), the number density evaluated beyond the dipole approximation can be much larger than the dipole-approximation values. In the next section, in order to gain a complete quantitative picture, we compute the total number of pairs.

\section{Total number of pairs}\label{sec:total}
In this section we discuss finally the total particle yield and compare the results obtained within the uniform-field approximation and beyond it. In particular, we perform the numerical integration of the density function $n(\boldsymbol{p})$:
\begin{equation}
\mathcal{N} = 2 \! \int \!\! n(\boldsymbol{p}) \mathrm{d}\boldsymbol{p}, \label{eq:n_total}
\end{equation}
where the factor $2$ appears due to the spin degeneracy. The total number of pairs $\mathcal{N}$ represents an extremely important characteristic which has a direct relation to the experiment and is a very useful indicator in comparison of various computational approaches. On the other hand, the calculation of this quantity is rather time consuming, especially beyond the dipole approximation where the cylindrical symmetry is broken by the appearance of the magnetic field. Nevertheless, we carry out the calculations for various values of the fast-pulse frequency $\omega$ (see Table~\ref{table:3}). We also evaluate the full enhancement factor $\mathcal{K}$ which is defined as $\mathcal{K}=\mathcal{N}(\text{I+II})/[\mathcal{N}(\text{I})+\mathcal{N}(\text{II})]$, where the particle yield in the case of the individual strong pulse is independent of $\omega$.
\begin{table}[h]
\centering
\setlength{\tabcolsep}{0.5em}
\begin{tabular}{c|crr|crr}
\hline \hline
\multirow{2}{*}{$\omega / m$}       & \multicolumn{3}{c|}{$\mathcal{N}$ (DA)}                                                                             & \multicolumn{3}{c}{$\mathcal{N}$ (bDA)}                                                  \\ \cline{2-7} 
& \multicolumn{1}{c|}{II} & \multicolumn{1}{c|}{I+II} & \multicolumn{1}{c|}{$\mathcal{K}$} & \multicolumn{1}{c|}{II} & \multicolumn{1}{c|}{I+II} & \multicolumn{1}{c}{$\mathcal{K}$} \\ \hline
\multicolumn{1}{r|}{$0.30$} & \multicolumn{1}{r|}{$<10^{-11}$} & \multicolumn{1}{r|}{$3.4\times 10^{-5}$}   & $5.1$                      & \multicolumn{1}{r|}{$<10^{-11}$} & \multicolumn{1}{r|}{$1.6\times 10^{-5}$}   & $2.5$       \\
\multicolumn{1}{r|}{$0.35$} & \multicolumn{1}{r|}{$<10^{-11}$} & \multicolumn{1}{r|}{$6.2\times 10^{-5}$}   & $9.3$                      & \multicolumn{1}{r|}{$<10^{-11}$} & \multicolumn{1}{r|}{$2.2\times 10^{-5}$}   & $3.3$        \\
\multicolumn{1}{r|}{$0.40$} & \multicolumn{1}{r|}{$3.5\times 10^{-11}$} & \multicolumn{1}{r|}{$1.1\times 10^{-4}$}   & $16$ & \multicolumn{1}{r|}{$1.1\times 10^{-11}$} & \multicolumn{1}{r|}{$3.3\times 10^{-5}$}   & $4.9$                      \\
\multicolumn{1}{r|}{$0.45$} & \multicolumn{1}{r|}{$1.8\times 10^{-9}$} & \multicolumn{1}{r|}{$2.0\times 10^{-4}$}   & $30$                      & \multicolumn{1}{r|}{$1.7\times 10^{-10}$} & \multicolumn{1}{r|}{$5.0\times 10^{-5}$}   & $7.5$      \\
\multicolumn{1}{r|}{$0.50$} & \multicolumn{1}{r|}{$9.3\times 10^{-10}$} & \multicolumn{1}{r|}{$3.5\times 10^{-4}$}   & $53$ & \multicolumn{1}{r|}{$2.4\times 10^{-10}$} & \multicolumn{1}{r|}{$7.8\times 10^{-5}$}   & $12$                    \\
\multicolumn{1}{r|}{$0.55$} & \multicolumn{1}{r|}{$6.4\times 10^{-7}$} & \multicolumn{1}{r|}{$6.9\times 10^{-4}$}   & $94$ & \multicolumn{1}{r|}{$4.3\times 10^{-8}$} & \multicolumn{1}{r|}{$1.3\times 10^{-4}$}   & $19$        \\
\multicolumn{1}{r|}{$0.60$} & \multicolumn{1}{r|}{$4.8\times 10^{-7}$} & \multicolumn{1}{r|}{$1.3\times 10^{-3}$}   & $180$                     & \multicolumn{1}{r|}{$6.4\times 10^{-8}$} & \multicolumn{1}{r|}{$2.1\times 10^{-4}$}   & $31$        \\
\multicolumn{1}{r|}{$0.65$} & \multicolumn{1}{r|}{$5.0\times 10^{-7}$} & \multicolumn{1}{r|}{$2.3\times 10^{-3}$}   & $320$                      & \multicolumn{1}{r|}{$8.1\times 10^{-8}$} & \multicolumn{1}{r|}{$3.5 \times 10^{-4}$}   & $51$        \\
\multicolumn{1}{r|}{$0.70$} & \multicolumn{1}{r|}{$1.8\times 10^{-4}$} & \multicolumn{1}{r|}{$4.1\times 10^{-3}$}   & $22$                      & \multicolumn{1}{r|}{$1.1\times 10^{-7}$} & \multicolumn{1}{r|}{$5.7\times 10^{-4}$}   & $84$        \\
\multicolumn{1}{r|}{$0.75$} & \multicolumn{1}{r|}{$1.8\times 10^{-4}$} & \multicolumn{1}{r|}{$7.2\times 10^{-3}$}   & $39$                      & \multicolumn{1}{r|}{$2.0\times 10^{-5}$} & \multicolumn{1}{r|}{$9.3\times 10^{-4}$}   & $35$        \\
\multicolumn{1}{r|}{$0.80$} & \multicolumn{1}{r|}{$2.5\times 10^{-4}$} & \multicolumn{1}{r|}{$1.2\times 10^{-2}$}   & $48$                      & \multicolumn{1}{r|}{$2.4\times 10^{-5}$} & \multicolumn{1}{r|}{$1.5\times 10^{-3}$}   & $48$        \\
\hline \hline            
\end{tabular}
\caption{The total number of pairs $\mathcal{N}$ produced in the presence of the individual fast pulse (II) and both the fast and the strong pulse (I+II) for various values of the fast-pulse frequency $\omega$. The results were obtained in the dipole approximation (DA) and beyond it (bDA). The values of $\mathcal{N}$ are displayed in units of $\lambdabar^{-3}$ where $\lambdabar$ is the reduced Compton wavelength of the electron ($\lambdabar \approx 386~\text{fm}$). The particle yield $\mathcal{N}(\text{I})$ amounts to $6.6 \times 10^{-6}~(\lambdabar^{-3})$.}
\label{table:3}
\end{table}
It is seen now that the dipole approximation indeed overestimates the amount of pairs. Our calculations confirm the other findings of Sec.~\ref{sec:enhancement}. Namely, one observes that the enhancement factor is almost insignificant for $\omega \lesssim 0.3 m$. Besides, the individual contribution of the weak pulse becomes larger than that of the strong pulse for $\omega \gtrsim 0.7 m$ (DA) and $\omega \gtrsim 0.8 m$ (bDA). Within the interval $0.3 m \lesssim \omega \lesssim 0.7 m$, the enhancement factor in the dipole approximation can reach a value of about $300$. However, according to the results obtained beyond this approximation, the total particle yields are about $1$ order of magnitude smaller.

\section{Discussion and conclusion}\label{sec:discussion}

Within the present investigation, we examined the main characteristics of the dynamically assisted Schwinger effect going beyond the previously used dipole approximation. In particular, we took into account the coordinate dependence of the fast weak pulse. It turned out that according to these more precise calculations, the patterns established in the homogeneous-field approximation cannot always be expected to provide the real features of the pair-production process. Instead, our results suggest that one has to take into account the spatial dependence of the external field in order to obtain more accurate quantitative and qualitative predictions.

\indent We summarize our main findings below:
\begin{itemize}
\item The structure of the momentum spectra of particles created becomes significantly different beyond the dipole approximation. The number of resonant peaks can double, and the momentum distributions along all three directions $x$, $y$, and $z$ become quite different.
\item Within the dipole approximation, the transversal momentum distribution never contains resonances corresponding to an even number of photons absorbed. However, beyond the dipole approximation, such peaks do appear unless the momentum along the propagation direction vanishes ($p_z = 0$).
\item The momentum spectra obtained in the dipole approximation and beyond it exhibit different quantitative behavior. While the latter mostly correspond to smaller values of the production probability, they can also have higher peaks. In order to accurately predict the quantitative characteristics of the spectra, one has to perform the calculations beyond the dipole approximation.
\item The enhancement of the particle yield due to the dynamical assistance, which is the essence of the processes considered in our study, turns out to be overestimated in the dipole approximation. The more precise calculations predict an enhancement factor that is several times smaller together with particle yields that are about $1$ order of magnitude smaller.
\end{itemize}

Although the external background considered in the present study incorporates the spatiotemporal dependence of the laser field, further steps towards studying more realistic configurations can also be taken. First, one can examine pulses of a finite size instead of two infinite pulses forming a standing wave. According to the recent studies~\cite{aleksandrov_prd_2017_2, lv_pra_2018}, the coordinate dependence of the envelope function can play a very important role, especially in the case of short laser pulses. Besides, in many studies of various scenarios within the dipole approximation, it was demonstrated that the momentum spectra of particles and other characteristics can be very sensitive to changes in the shape of the laser pulse (see, e.g., Refs.~\cite{aleksandrov_prd_2017_1, linder_prd_2015, abdukerim_plb_2013, otto_epja_2018, abdukerim_cpb_2017}). The analysis of the pulse shape effects beyond the dipole approximation is an important issue to be investigated further.

Finally, we stress that the spatial dependence of the external background in the context of Schwinger pair production was considered so far in a very few studies~\cite{aleksandrov_prd_2016, aleksandrov_prd_2017_2, schneider_jhep_2016, torgrimsson_arxiv_dec_2017, lv_pra_2018, kohlfuerst_prd_2018}. We expect that multidimensional inhomogeneities should be significant 
for a much broader class of possible scenarios.

\section*{Acknowledgments}
This study was supported by RFBR-DFG (Grants No.~17-52-12049 and No. PL 254/10-1) and by SPbSU-DFG (Grants No.~11.65.41.2017 and No.~STO 346/5-1). I. A. A. also acknowledges the support from the German-Russian Interdisciplinary Science Center (G-RISC) funded
by the German Federal Foreign Office via the German Academic Exchange Service (DAAD), from TU Dresden (DAAD-Programm  Ostpartnerschaften), from the FAIR-Russia Research Center, and from the Foundation for the advancement of theoretical physics and mathematics ``BASIS.''


\end{document}